\documentclass{pazha2}
\usepackage{graphicx}
\usepackage{multirow}
\usepackage{color}
\definecolor{darkblue}{rgb}{0,0,0.9}
\newcommand\myfontsizt{\fontsize{15.8pt}{13pt}\selectfont}
\newcommand\myfontsiza{\fontsize{9.6pt}{13pt}\selectfont}

\def\*{$^{*}$}
\def\a{$^{\mbox{\small a}}$}
\def\bb{$^{\mbox{\small b}}$}
\def\v{$^{\mbox{\small c}}$}
\def\g{$^{\mbox{\small d}}$}
\def\dd{$^{\mbox{\small e}}$}
\def\e{$^{\mbox{\small f}}$}
\def\zg{$^{\mbox{\small g}}$}

\def\etal{{et~al.}}
\begin{document}
\journalinfo{to be}{2024}{50}{1}{1}{3}{27}[21]
\sloppypar

\title{\myfontsizt\bf GRB\,231115A - a magnetar giant flare in the M82 galaxy}
\year=2024

\author{\myfontsiza\bf P.~Yu.~Minaev\email{\normalsize minaevp@mail.ru}\address{1,2}, 
A.~S.~Pozanenko\address{1,3}, 
S.~A.~Grebenev\address{1},
I.~V.~Chelovekov\address{1}, \protect\\ 
N.~S.~ Pankov\address{1,3},
A.~A.~Khabibullin\address{3},
R.~Ya.~Inasaridze\address{4},
A.~O.~Novichonok\address{5}
\addrestextf{1}{Space Research Institute, Russian Academy of Sciences, Profsoyuznaya ul. 84/32, Moscow 117997, Russia}
\addrestextl{2}{P.N. Lebedev Physical Institute, Russian
  Academy of Sciences, Leninskiy pr. 53, Moscow 119991, Russia}
\addrestextl{3}{National Research University ''Higher School of
  Economics'', Myasnitskaya ul. 20, Moscow 101000, Russia}
\addrestextl{4}{Evgeni Kharadze Georgian National Astrophysical Observatory, Mount Kanobili, Abastumani, Georgia}
\addrestextl{5}{Petrozavodsk State University, Lenina ul. 33, 
Petrozavodsk 185910, Karelia, Russia}
}

\shortauthor{MINAEV ET AL.}
\shorttitle{GRB 231115A -- A MAGNETAR GIANT FLARE}

\submitted{18.11.2023}
\revised{20.11.2023}
\accepted{21.11.2023}

\begin{abstract}
\noindent
The results of a study of the short gamma-ray burst GRB~231115A
in the X-ray and gamma-ray ranges are presented, based on data
from the INTEGRAL and {\sl Fermi\/} space observatories. The
source of the burst is localized by the IBIS/ISGRI telescope of
INTEGRAL observatory with an accuracy of $\leq1.5\arcmin$, it is
located in the Cigar Galaxy (M\,82). Early follow-up
observations of the burst localization region were carried out
in the optical range with the 36-cm telescope of the ISON-Kitab
observatory and the 70-cm telescope AS-32 of the Abastumani
Astrophysical Observatory, the optical emission has not been
detected. The proximity of the host galaxy ($D_L \simeq 3.5$
Mpc) significantly limits energetics of the event
($E_{iso}\ \sim\ 10^{45}$ erg) and allows us to interpret the
burst as a giant flare of a previously unknown soft gamma
repeater (SGR) which is an extreme manifestation of the activity
of a highly magnetized neutron star (magnetar). This conclusion
is confirmed by the energy spectrum atypically hard for cosmological
gamma-ray bursts, as well as the absence of optical afterglow
and gravitational wave signal, which should have been detected
in the LIGO/Virgo/KAGRA experiments if the burst was caused by a
merger of binary neutron stars. The location of the burst in the
$E_{p,i}$ -- $E_{iso}$ and $T_{90,i}$ -- $EH$ diagrams also
suggests that GRB~231115A was a magnetar giant flare. This is
the first well-localized giant flare of an extragalactic SGR.

\end{abstract}

\section{INTRODUCTION}
\noindent
The presence of two different types of gamma-ray bursts (GRBs)
was discovered in the KONUS experiment (Mazetz et al. 1981) and
later confirmed by the CGRO/BATSE detector (Kouveliotou et
al. 1993) when analyzing the distribution of bursts by the
duration parameter $T_{90}$. Short bursts (lasting less than
2~s) are characterized by a harder energy spectrum (with a
larger portion of high-energy emission) and a less pronounced
spectral evolution (lag of low-energy emission relative to
high-energy one) compared to long bursts ($T_{90}\ga 2$ s,
e.g. Kouveliotou et al. 1993; Norris et al. 2005; Minaev et
al. 2010a, 2012, 2014). At the same time, duration and spectral
hardness distributions for these two types traditionally used
for classification of gamma-ray bursts, overlap significantly,
leaving the problem of burst classification in the intersection
area of distributions relevant up to the present day (see, e.g.,
Dezalay et al.  1997; Minaev et al. 2010b; Minaev \& Pozanenko
2017; Tarnopolsky 2019).

It is believed that the short gamma-ray bursts (later designated
as type I bursts) are associated with a merger of two neutron
stars (Blinnikov et al. 1984; Paczynski 1986; Meszaros \& Rees
1992), which has been recently confirmed by the detection of
the GRB/GW 170817 and GRB/GW 190425 events by the LIGO/Virgo
gravitational wave detectors (Abbott et al. 2017a,b; Pozanenko
et al. 2018, 2019). Some type I bursts are accompanied by an
additional emission component with a duration of tens of seconds
and a softer (compared to the main episode of emission) spectrum
--- extended emission, which nature has not yet been clarified
(Connaughton 2002; Gehrels et al. 2006; Rosswog 2007; Metzger et
al. 2008; Minaev et al. 2010a; Norris et al. 2010; Barkov \&
Pozanenko 2011).

The long (type II) gamma-ray bursts are associated with the core
collapse of a massive star (Woosley 1993; Paczynski 1998;
Meszaros 2006), some of them, closest to the observer, are
accompanied by type Ic supernovae (see, e.g., Galama et al.
1998; Paczynski 1998; Cano et al. 2017; Volnova et al. 2017;
Belkin et al. 2020, 2023).

There are anomalies in the correlation between the duration of
gamma-ray bursts and their type when some short gamma-ray bursts
were accompanied by supernova explosions (e.g., GRB 200826A,
Rossi et al. 2022) or, conversely, the long GRB 230307A was
associated with a kilonova (Levan et al. 2023). Thus, the
correct classification of gamma-ray bursts, along with
determining the redshift of their host galaxies, is important
for studying their sources.

Short bursts of hard emission are also characteristic of some
soft gamma repeaters (SGR, Golenetskij et al. 1979; Mazets et
al. 1979a) during their extreme activity (the so-called giant
flares, e.g. Mazets et al. 1979b, 2008; Thompson \& Duncan 2001;
Frederiks et al. 2007). The light curve of a giant flare
consists of a short (fractions of a second), hard and very
bright main episode, which may be followed by a long (hundreds
of seconds) and much weaker extended emission, characterized by
periodicity connected with rotation of a neutron star in which
magnetosphere the giant flare occurred.

All soft gamma repeaters confirmed by long-term observations are
located in the Galaxy, so far, giant flares have been detected
from four of them. However, the main short episode of a giant
flare could also be detected from some nearby galaxies. For
example, a giant flare from SGR~1806-20 could be confidently
detected at a distance of 30--50 Mpc (Hurley et al. 2005; Nakar
et al. 2005). Several candidates to giant SGR flares, possibly
originating in nearby galaxies, have been proposed, based on the
IPN triangulation results (see, e.g., Frederiks et al. 2007;
Mazets et al. 2008).

Observed properties of the giant flares (time profile, hardness
and spectral evolution of the emission) and their frequency (no
repeated giant flares have been observed from any known SGR so
far) are largely similar to the properties of type I GRBs. This
introduces some difficulty in classification of transient
gamma-ray events (Mazets et al. 2008; Minaev \& Pozanenko
2020b). The most reliable method for identifying SGR sources is
the detection of periodicity in a tail of their light
curves. Periodicity has been found for many galactic SGRs, for
example: SGR~0520-66 (Mazets et al. 1979b), SGR~1806-20 (Mazets
et al. 2005; Palmer et al.  2005), SGR~1900+14 (Mazets et
al. 1999; Feroci et al. 1999). Periodicity was found after a few
short gamma-ray bursts detected in the BATSE/CGRO experiment,
for example: GRB 930905 (Pozanenko et al. 2005) and GRB 970110
(Crider 2006), and these gamma-ray bursts also may be considered
as candidates to giant flares of unidentified SGRs.

Soft gamma repeaters are most likely connected with magnetars
--- single neutron stars with extremely strong magnetic fields
(${B} \ga 10^{14}$ G). The incredible power and physical origin
of their giant flares remain unclear (Duncan \& Thompson 1992;
Thompson \& Duncan 1995; Kouveliotou et al. 1999).

Localization of the short gamma-ray burst GRB 231115A with an
accuracy better than 2\arcmin, carried out as part of the
operational (Quick Look) analysis of telemetry data from the
IBIS/ISGRI gamma-ray telescope of the INTEGRAL observatory,
reliably connects it with the nearby galaxy M\,82 with active
star formation (Burns 2023), which allows one to suggest this
event being caused not by the merger of a pair of neutron stars,
but by a giant flare of a previously unknown soft gamma repeater
located in this galaxy (see, for example, D'Avanzo et al. 2023a;
Minaev \& Pozanenko 2023b).

The paper presents the results of observations and analysis of
GRB 231115A in gamma-rays based on data from INTEGRAL and {\sl
  Fermi\/} observatories and our own early optical observations
of the localization region in order to clarify the nature of the
burst source. In particular, we used the classification based on
the correlation of the total energy parameter $E_{iso}$,
spectral hardness $E_{p,i}$ (Amati et al. 2002), and duration of
gamma-ray bursts in the rest frame $T_{90,i}$, proposed by
Minaev \& Pozanenko (2020a,b).

\section*{GRB 231115A DETECTION AND EARLY OBSERVATIONS}
\noindent
GRB 231115A with a duration of about 0.1~s and a hard emission
spectrum, which is characteristic of both type I (short)
gamma-ray bursts and giant flares of magnetars (SGRs), was
detected on November 15, 2023 at 15\uh36\um21\fs20 UT by
following space gamma-ray detectors: {\sl Fermi}/GBM (Dalessi et
al. 2023), INTEG\-RAL/IBIS/ISGRI (Mereghetti et al. 2023),
KONUS-Wind (Frederiks et al. 2023), Glowbug (Cheung et
al. 2023), Insight-HXMT/HE (Xue et al. 2023), and Swift/BAT
(Roncini et al.  2023).

The burst was detected in the field of view of the IBIS/ISGRI
telescope, therefore its position was determined with an
accuracy better than 2\arcmin, which made it possible to
identify the host galaxy --- M\,82 (D'Avanzo et al. 2023a;
Burns 2023; Mereghetti et al. 2023). The high localization
accuracy initiated a search for possible emission in other
energy ranges. Attempts to observe the optical afterglow
undertaken by many scientific groups at different telescopes
were unsuccessful (Lipunov et al. 2023a,b; Balanutsa et
al. 2023; Iskandar et al. 2023; Chen et al. 2023; Jiang et
al. 2023; Hayatsu et al. 2023; Perley et al. 2023; D'Avanzo et
al. 2023b; Turpin et al. 2023; An et al. 2023; Hu et
al. 2023). An optical candidate discovered by the 0.7-m
GROWTH-India telescope (Kumar et al. 2023a) was found to be an
artifact after more in-depth analysis (Kumar et al. 2023b).

The Swift/XRT and NuSTAR X-ray telescopes detected no signs of
an X-ray afterglow 2.5 and 4 hours after the trigger, respectively
(Osborne et al. 2023; Grefenstette \& Brightman 2023). The MAGIC
experiment obtained an upper limit on the gamma-ray flux in the
range above 250 GeV 8 hours after the burst (MAGIC
Collaboration, 2023). The CHIME/FRB radio telescope has also not
detected any activity from the source, similar to a fast radio
burst (FRB) in the range of 400--800 MHz (Curtin 2023). 

Finally, the LIGO/Virgo/KAGRA gravitational-wave detectors have
not detected the signal that should have accompanied the merger
of two neutron stars if the detected burst was indeed a short
gamma-ray burst caused by such mergers (LIGO Collaboration,
2023). A neutrino signal has also not been detected in the
IceCube experiment (IceCube Collaboration, 2023).
\begin{figure}[!t]
    \centering
    \includegraphics[width=1.05\linewidth]{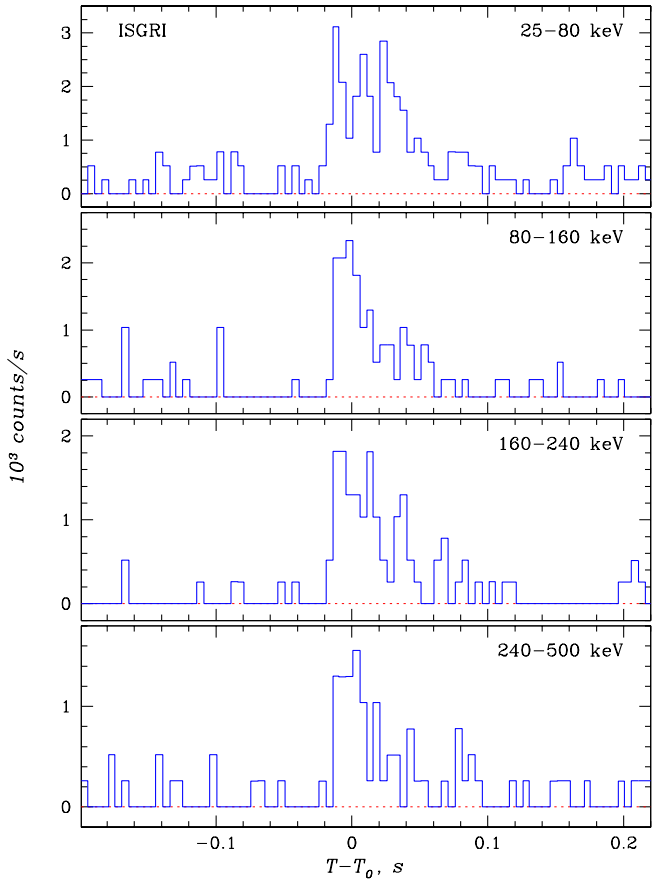}
    \caption{Time profile of GRB\,231115A according to data from
      the IBIS/ISGRI telescope of the INTEGRAL observatory in
      four energy bands with a resolution of 5 ms.}
    \label{fig:ibis-lcurve}
\end{figure}

\section*{DATA ANALYSIS OF THE INTEGRAL OBSERVATORY}
\noindent
The INTErnational Gamma-Ray Astrophysics Laboratory INTEGRAL
(Winkler et al. 2003; Kuulkers et al. 2021) has been operating
in a high apogee orbit for the 22nd year. There are several
wide-field telescopes on board with a coding aperture capable to
obtain sky images and perform a comprehensive analysis of the
energy spectra and variability of various cosmic sources: the
IBIS gamma-ray telescope with two detectors: ISGRI (Lebrun et
al.  2003), sensitive in the range of 20--400 keV, and PICsIT
(Labanti et al. 2003), sensitive in the range of 200 keV -- 10
MeV, the gamma-ray spectrometer SPI (Vedrenne et al. 2003) with
cooled germanium detectors, sensitive in the range of 20 keV --
8 MeV, and two JEM-X X-ray telescopes (Lund et al. 2003),
sensitive in the range of 4--30 keV. Note also the
anti-coincidence shield ACS of the SPI gamma-ray spectrometer
(Rau et al. 2005), which operates as a large-area
omnidirectional detector in the range 85 keV -- 10 MeV and
records the photon count rate with a time resolution of 50~ms.

The burst was detected during the scheduled observations of the
M\,81 galaxy field, carried out under the INTEGRAL AO-20
proposal \#\,2020020 (PI I.A. Mereminsky). Immediately after the
burst detection the data from all instruments in the time
interval, starting from $3$ hours before the burst and up to
$24$ hours after it, were transferred to the authors of this
work within the framework of the AO-20 proposal \#\,2040014 (PI
P.Yu. Minaev) for detailed analysis and comprehensive study of
the burst.
\begin{figure}[t]
\includegraphics[width=\linewidth]{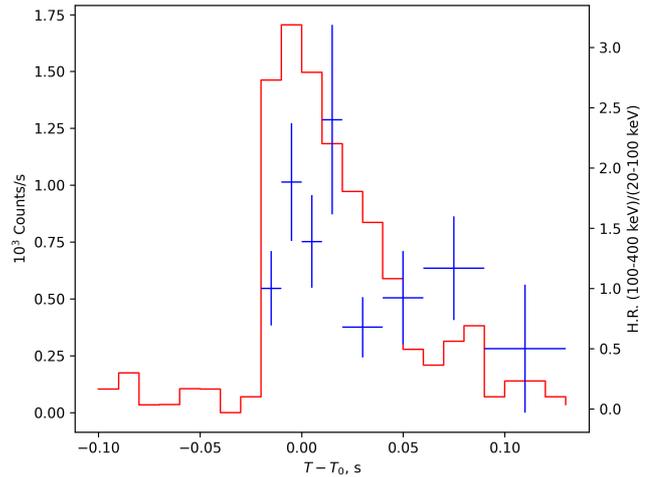}
\caption{Evolution of the hardness ratio of the GRB~231115A
  emission, the ratio of the number of photons recorded by
  the IBIS/ISGRI telescope in the bands 100--400 and 20--100 keV
  (blue crosses, scale on the right). For comparison, the red
  line shows the profile of the burst with a resolution of 10~ms
  in the wide energy range of 20--400 keV (normalization is
  shown on the left scale). In the first $\sim40$ ms of the
  burst, the hardness of its emission increased and was on
  average 1.5--2 times higher than in the subsequent $\sim 80$
  ms.}\label{fig:ISGRI_Hardness_Ratio}
\end{figure}

\subsection*{Time profile of the burst}
\noindent
The burst GRB\,231115A fell into the field of view of the main
telescopes of the observatory, which allowed them to detect and
study it nearly in real time using the automatic IBAS system
(Mereghetti et al. 2023). The system performs the Quick Look
analysis of data from the IBIS/ISGRI telescope and distributes
alerts on the localized gamma-ray bursts through the Gamma-ray
burst Coordinate Network (GCN) system.
\begin{figure}[!t]
\centering
\hspace{-2mm}\includegraphics[width=1.05\linewidth]{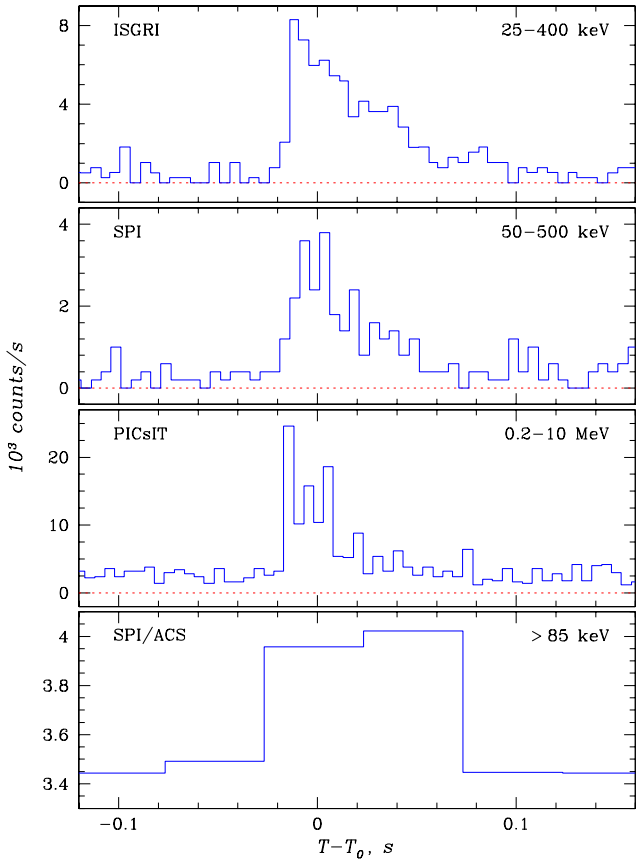}
\caption{Time profile of GRB\,231115A, based on data from four
  instruments of the INTEGRAL observatory: the IBIS/ISGRI
  (25--400 keV) and IBIS/PICsIT (0.2--10 MeV) telescopes, the
  SPI gamma-ray spectrometer (20--500 keV) and its ACS
  shield ($>85$ keV). Temporal resolution is 5~ms, except for the
  profile obtained by the SPI-ACS detector, having a maximum
  resolution of 50~ms.}\label{fig:igral-lcurves}
\end{figure}

Figure~\ref{fig:ibis-lcurve} shows the time profiles of the
burst obtained by the IBIS/ISGRI detector in four different
energy bands (count rate records with a resolution of 5 ms). The
time is counted from the trigger time of the event by the {\sl
  Fermi}/GBM monitor (Dalessi et al. 2023), corrected for 0.47 s
of a time delay due to the large distance of the INTEGRAL
spacecraft from the Earth at the moment of the burst. It is seen
that the burst is hard (clearly observed up to 500 keV) and has
a FRED-like (fast rise --- exponential decay) time profile with
a total duration of less than 120 ms at energies $\ga 80$
keV. This allows us to classify it as a type I (short) burst or
a giant flare of a previously unknown magnetar. At the same
time, it is obvious that the maximum number of photons was
recorded in the softest IBIS/ISGRI band ($E\la 80$ keV). The
profile of the burst changes in this band and becomes
pedestal-shaped with a wide ($\Delta T\simeq 60$ ms) top,
possibly behind counting the superposition of several short
flashes at once.

This is also confirmed by the evolution of hardness of the
GRB~231115A emission shown in
Fig.~\ref{fig:ISGRI_Hardness_Ratio}. The emission hardness (blue
crosses, right scale) is defined as the ratio of the number of
photons detected by the IBIS/ISGRI telescope in the 100--400 and
20--100 keV bands. For comparison, the red line shows the time
profile of the burst in the wide 20--400 keV range (only photons
were taken into account whose probability of association with
the burst exceeded 20\%). It is seen that the hardest emission
was recorded during the first $\sim 40$ ms of the burst, and all
this time the hardness gradually increased. Its average value
$HR=1.55\pm0.22$. During the next $\sim 80$ ms, the hardness
ratio significantly (at the level of $3.9\sigma$) decreased by
1.9 times to the value $HR=0.80\pm0.17$. A drop in hardness
ratio towards the end of the event is observed in many gamma-ray
bursts. In giant flares of magnetars, the evolution of the
hardness ratio is poorly understood due to their exceptional
brightness, which saturates the telemetry of most instruments.

The burst has been also detected by other instruments on board
INTEGRAL. In Fig.~\ref{fig:igral-lcurves} its time profiles
obtained from data of the gamma-ray telescopes SPI, IBIS/PICsIT,
and SPI-ACS are presented in comparison with the broad-band
profile obtained by the IBIS/ISGRI telescope in the range of
25--400 keV. It is seen that the burst is confidently detected
up to 500 keV and above. The profiles in many aspects (up to
statistical errors) repeat each other. Note that the slightly
skewed shape of the burst profile measured by the SPI/ACS
detector can be explained by the fact that the first time bin in
the photon count rate record during the burst began noticeably
before the burst, which lowered the total number of photons
measured in the bin. The source was not detected by the
\mbox{JEM-X} telescopes, although it was only $3\fdg8$ from the
center of their field-of-view in the region of their fairly high
sensitivity. The upper limit on the X-ray flux from the burst in
the 3--20 keV range (at the $1\sigma$ level), obtained from data
of both the JEM-X telescopes assuming that the event had a
duration of 50 ms in this energy range, is 1.9 Crab, which
corresponds to the flux $4.7\times10^{-8}\ \mbox{erg
  s}^{-1}\ \mbox{\rm cm}^{-2}$.
\begin{figure*}
\centering
\includegraphics[width=0.9\linewidth]{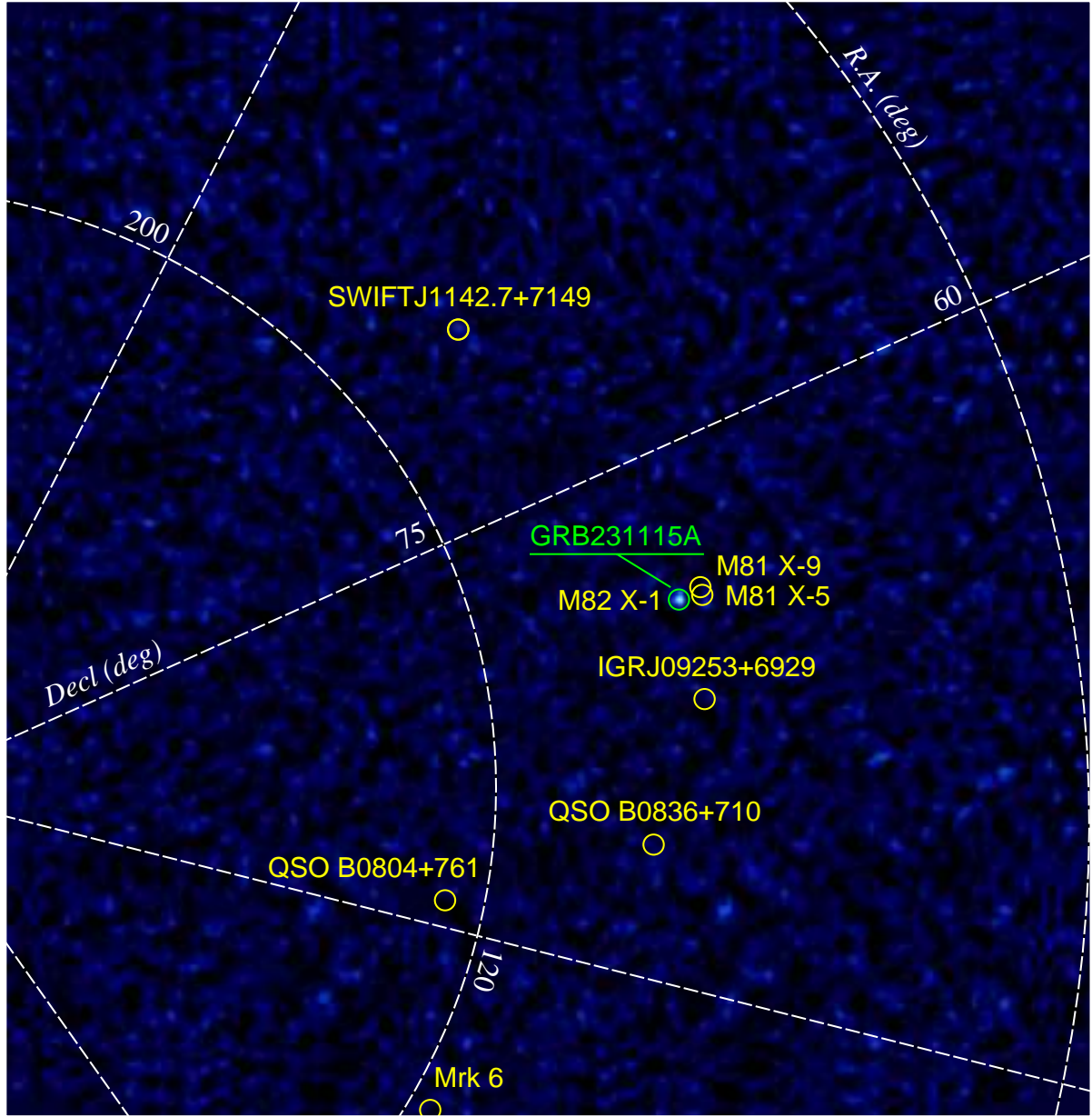}
\caption{X-ray image of the sky (the $S/N$ ratio map) within the
  field of view of the IBIS/ISGRI telescope of the INTEGRAL
  observatory, obtained during the gamma-ray burst
  GRB\,231115A. The size is $29\deg\times29\deg$, the exposure
  is 120 ms, the energy range is 20--400 keV. The known persistent 
  sources are indicated in the field. GRB\,231115A, the only
  source confidently detected (at $S/N\simeq10.5$), coincides in
  location with the M\,82 galaxy (on the map --- with the position
  of the ultra-luminous X-ray source M\,82 X-1, located at its
  central region).}\label{fig:ibis-bigmap}
\end{figure*}
\begin{figure*}
\centering
\includegraphics[width=0.8\linewidth]{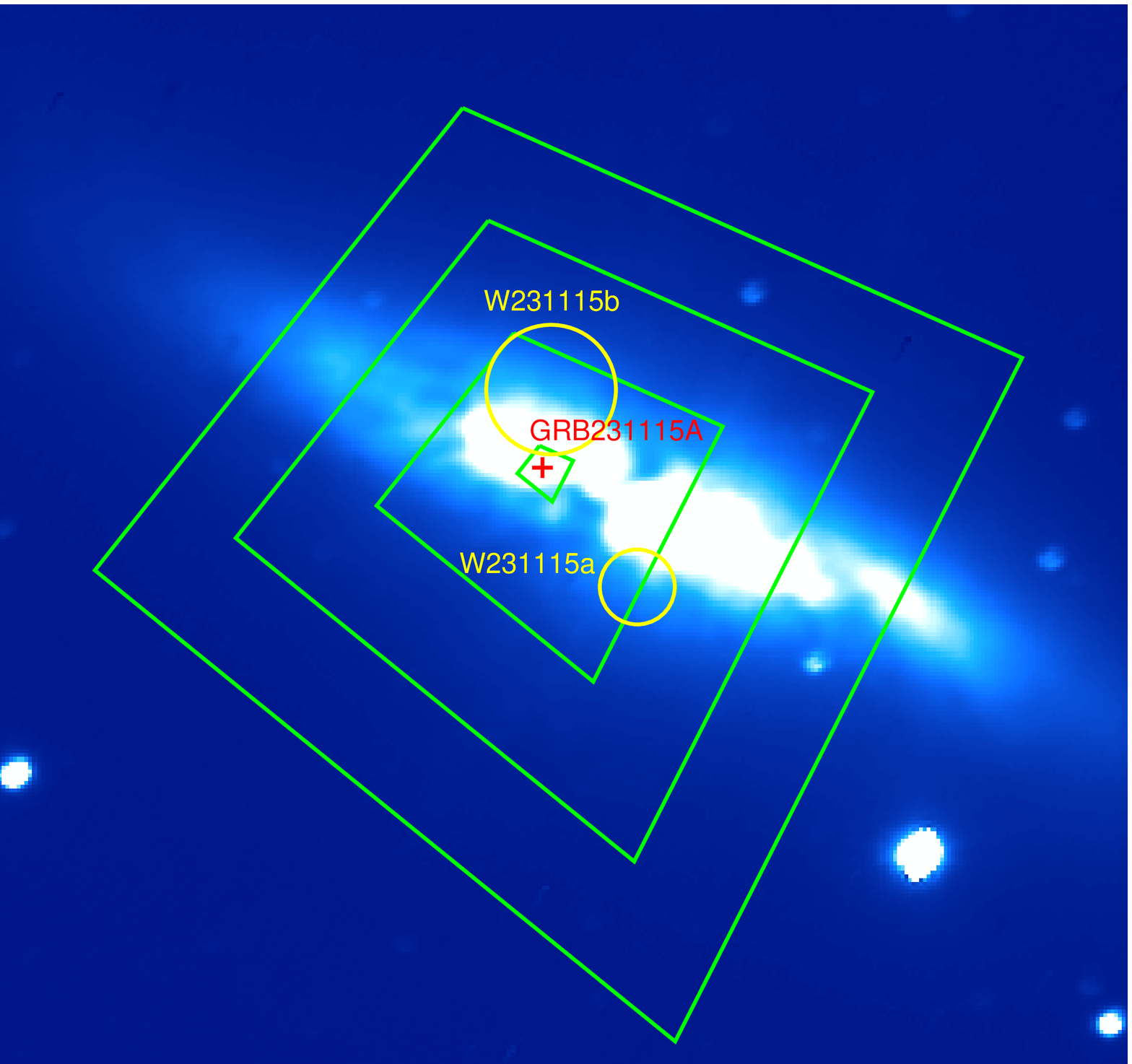}
\caption{Image of the GRB\,231115A localization region obtained
  with the 70-cm AS-32 telescope of the Abastumani Astrophysical
  Observatory 10.7 hours after the burst. Green diamonds show
  the burst localization contours on the X-ray map of the $S/N$
  ratio obtained by the INTEGRAL IBIS/ISGRI telescope, the third
  one from the center corresponds to the 90\% confidence level
  ($1\farcm5$ uncertainty). Circles represent two proposed
  optical candidates (Hu et al. 2023, see text). It is seen that
  the source of the burst is most likely located in the bright
  disk of the M\,82 galaxy, where it is difficult to observe,
  while the proposed candidates were found in the regions of
  smaller brightness at the edge of the localization region.}
    \label{fig:ibis-smallmap}
\end{figure*}
 
\subsection*{Localization}
\noindent
One of the most important results is the localization of
GRB\,231115A. The first information on coordinates of the new
burst was promptly distributed by the automatic IBAS system
(d'Avanzo et al. 2023a; Burns 2023; Mereghetti et al. 2023),
which allowed its observations to be quickly begun with optical
and radio telescopes over the
world. Figure~\ref{fig:ibis-bigmap} shows an image of the sky
within the field of view of the IBIS/ISGRI telescope of
$29\deg\times29\deg$ in size, accumulated over the duration of
the burst (120 ms) in the energy range of 20--400 keV. In such a
short time, only the burst itself was detected in the field at a
significant level (the signal-to-noise ratio $S/N\simeq 10.5$).

But the main result is that the figure clearly shows that the
burst occurred in the nearby well-known galaxy M\,82 (Cigar),
located at a distance of $D_L=3.5$ Mpc. The source coordinates
are
R.A.$=09\uh55\um59\fs28,$\ Decl.$=+69\deg41\arcmin02\farcs40$
($148\fdg997$, $+69\fdg684$; epoch 2000.0, uncertainty is
$1\farcm5$). Figure~\ref{fig:ibis-smallmap} shows the contours
corresponding to different levels of the $S/N$ ratio for the
source of the burst in this image, superimposed on the optical
(in the R filter) image of the M\,82 galaxy, obtained by us with
the 70-cm AS--32 telescope of the Abastumani Astrophysical
Observatory (10.7 hours after the burst). It is seen that the
center of localization lies in the disk of the galaxy, leaving
no doubt that it is the host of GRB\,231115A. There were
several other known extragalactic X-ray sources within the
telescope's field of view, including the ultraluminous sources in
the galaxies M\,81 and M\,82; none of them appeared in the image
during such a short exposure and, obviously, could not
contribute to the time profile of the burst.
\begin{figure}[!t]
\centering
\includegraphics[width=1.0\linewidth]{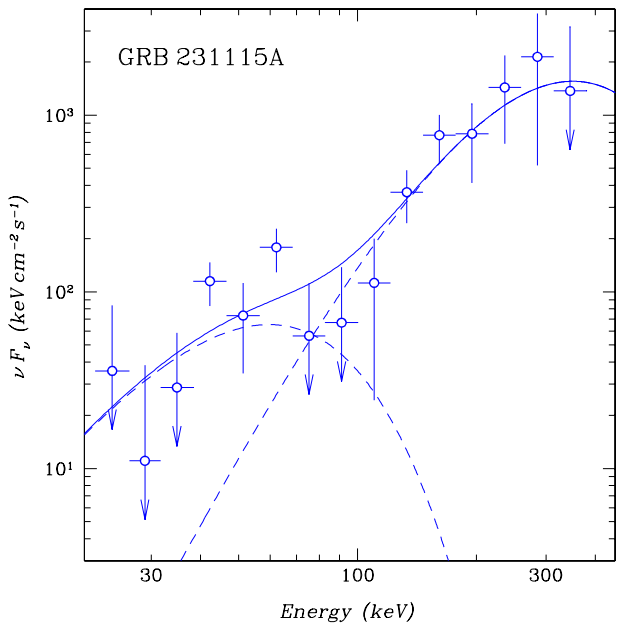}
\caption{Emission spectrum $\nu F_{\nu}$ of GRB\,231115A in the
  20--400 keV range according to data from the IBIS/ISGRI
  telescope of the INTEGRAL observatory averaged over the total
  ($\simeq 120$ ms) duration of the burst. The solid line shows
  the result of the best approximation with the {\sc BBR$+$CPL}
  model, the dashed lines show individual components of the
  model.}\label{fig:ibis-spec}
\end{figure}

\subsection*{Emission spectrum}
\noindent
Figure~\ref{fig:ibis-spec} shows the emission spectrum
$\nu\,F_{\nu}(\nu)$ of the gamma-ray burst GRB\,231115A,
obtained from the data of the IBIS/ISGRI telescope in the
20--400 keV range. The spectrum has been accumulated over the
entire duration of the event. It is seen that it is very hard,
the energy contained in the emission increases with increasing
the photon energy. An attempt to approximate the spectrum using
simple one-component models: {\sc powerlaw} (PL), {\sc cutoffpl}
(CPL) or {\sc bbodyrad} (BBR), has been not very successful,
which confirms the visual impression on the presence of two
components in the spectrum -- the soft and hard ones. The XSPEC
package developed at NASA/HEASARC (Arnaud et al. 1996) was used
for approximation. The results of approximation (best-fit
parameters and fluxes) are summarized in
Table~\ref{tab:ibis-spec}. The CPL model gives a slightly better
approximation:
\begin{equation}\label{eq:cpl}
I_{\nu}=A E^{-\alpha}\exp{(-E/E_c)},
\end{equation}
where $I_{\nu}$ is the photon spectrum, $\alpha$ is the photon
index, and $E_c$ is the characteristic energy of the high-energy
cutoff. Note that the average photon index for the burst
(according to the PL model) is $\alpha\simeq 0.46$, which means
that the spectral index ($=\alpha-1$) is negative and the
spectral density of emission increases with energy. Note also
that the exponential cutoff in the spectrum at high energies
according to the IBIS/ISGRI telescope data is, in fact, not very
reliable. In particular, the cutoff energy $E_{c}\simeq 330$
keV, obtained by approximating the spectrum with the CPL model,
was determined with large errors ($\sim73$ keV) and therefore
was then fixed and was not used as a free parameter. Thanks to
exact knowledge of the burst's host galaxy (M\,82) and the
luminosity distance to it ($D_L=3.5$ Mpc), it is possible to
determine the average luminosity of the burst in the range of
20--400 keV as $L_{X}\simeq 2.9\times10^{45}\ \mbox{\rm erg
  s}^{-1}$.

The burst spectrum may be more successfully approximated by the
two-component models BBR$+$BBR and BBR$+$CPL.  As can be seen
from Table~\ref{tab:ibis-spec}, in both cases the soft emission
component can be described by the black-body spectrum with the
temperature $kT_{bb}\sim 15$ keV and radius of the emitting
surface $R_{bb}\sim 170$ km. Note that the value of $R_{\rm bb}$
exceeds the typical radius $R_{ns}\simeq 12$ km of a neutron
star as well as the temperature $kT_{bb}$ exceeds by nearly an
order of magnitude the Eddington temperature for a neutron star
(if we actually observed a magnetar flare), thus the black-body
photosphere with such a temperature should expand and outflow
effectively. It is obvious that the use of the black-body
spectrum to describe the soft component of the burst emission
was not physically justified but was applied for simplicity and
convenience. If this component was indeed connected with the
expanded photosphere of a neutron star, its spectrum should have
been formed in result of Comptonization and had a Wien
shape. Note that the spectral index of the hard power-law
component of the emission (in the CPL model) is negative, and
therefore the emission spectrum rises very steeply towards high
energies. The normalization $\nu F_{\nu}(\nu)$ used for the
figure shows immediately that the main emission energy is in
the hardest part of the spectrum (it is contained in photons
with energies $h\nu\ga 300$ keV).
\begin{table*}[t]
\centering
\caption{Results of the analysis of the time-integrated spectrum
  of GRB 231115A in the 20--400 keV range based on the IBIS/ISGRI data}
\label{tab:ibis-spec}
\vspace{5mm}
\begin{tabular}{c|c|r@{$\pm$}l|r@{$\pm$}l|r@{$\pm$}l|c} \hline\hline
Model & $\chi^2/N$\a\ &\multicolumn{2}{c|}{A}&\multicolumn{2}{c|}{$\alpha$\bb}&
\multicolumn{2}{c|}{$kT_{bb},\ E_{c},$\dd} & Flux\e, \\	
    &
&\multicolumn{2}{c|}{}&\multicolumn{2}{c|}{}&\multicolumn{2}{c|}{keV}&$10^{-6}$
erg s$^{-1}$ cm$^{-2}$  \\ \hline
&&\multicolumn{2}{c|}{}&\multicolumn{2}{c|}{}&\multicolumn{2}{c|}{}&\\ [-3.0mm]
PL  &30.6/29&$1.97$&$0.26$\v\ &$0.46$&$0.03$&\multicolumn{2}{c|}{--}& $1.72\pm0.23$\\
BBR &31.1/29&$22.2$&$1.48$\g\ &\multicolumn{2}{c|}{--}&$91$&$5$ &$1.99\pm0.27$\\
CPL &29.8/28&$2.91$&$0.38$\v\ &$-0.07$&$0.03$&\multicolumn{2}{c|}{328}&$1.92\pm0.24$\\ \hline
&&\multicolumn{2}{c|}{}&\multicolumn{2}{c|}{}&\multicolumn{2}{c|}{}&\\ [-3.0mm]
BBR &26.6/27&$17.6$&$1.3$\g\  &\multicolumn{2}{c|}{--}&$106$&$7$   &   $1.92\pm0.29$\\ [-1.3mm]
$+$ &&\multicolumn{2}{c|}{}&\multicolumn{2}{c|}{}&\multicolumn{2}{c|}{}&\\ [-1.1mm] 
BBR  &       & $170$&$35$\g\    &\multicolumn{2}{c|}{--}&$14.1$&$1.6$&
$0.09\pm0.04$\\  [1mm] \hline
&&\multicolumn{2}{c|}{}&\multicolumn{2}{c|}{}&\multicolumn{2}{c|}{}&\\ [-3.0mm]
CPL&24.9/26&$4.91$&$0.77$\v\ &$-2.50$&$0.03$         &$78$ &$5$   &$2.01\pm0.31$\\ [-1.3mm]
$+$ &&\multicolumn{2}{c|}{}&\multicolumn{2}{c|}{}&\multicolumn{2}{c|}{}&\\ [-1.1mm] 
BBR&       &$178$&$26$\g\    &\multicolumn{2}{c|}{--}&$15.2$&$1.3$&  $0.13\pm0.04$  \\ \hline
\multicolumn{9}{c}{}\\ [-2mm]
\multicolumn{9}{l}{\a\ The minimum value of $\chi^2$ and the number of degrees of freedom $N$.}\\
\multicolumn{9}{l}{\bb\ Photon index of the power-law component $I_{100}\ (E/100\ \mbox{\rm keV})^{-\alpha}$.}\\
\multicolumn{9}{l}{\v\ Normalization of this component $I_{100}$ at 100 keV [$10^{-2}\ \mbox{phot s}^{-1}$ cm$^{-2}$ keV$^{-1}$].}\\
\multicolumn{9}{l}{\g\ Radius of the radiating surface
  $R_{bb}$ [km]  at the distance $D_{L}=3.5$ Mpc.}\\ 
\multicolumn{9}{l}{\dd\ Temperature $kT_{bb}$ or exponential cutoff energy $E_{c}$.}\\
\multicolumn{9}{l}{\e\ Flux in the 20--400 keV range.}\\
\end{tabular}
\end{table*}

\section*{{\sl Fermi}/GBM DATA ANALYSIS} 
\noindent
The monitor of gamma-ray bursts (GBM) on board the {\sl Fermi\/}
observatory consists of 12 NaI scintillaton detectors, sensitive
in the 8--1000 keV range, and 2 BGO scintillaton detectors,
sensitive in the 0.2--40 MeV range. It is designated for the
registration and comprehensive study of cosmic gamma-ray bursts
(Meegan et al. 2009; Paciesas et al. 2012).  The source of data
from the {\sl Fermi}/GBM monitor in this study was a public FTP
archive {\sl legacy.gsfc.nasa.gov/fermi/data\/}. Again the
moment of the burst trigger by {\sl Fermi}/GBM, November 15,
2023 at 15\uh36\um21\fs20 UT, is used as zero on a time scale.
\begin{figure}[!t]
\centering
\includegraphics[width=1.06\columnwidth]{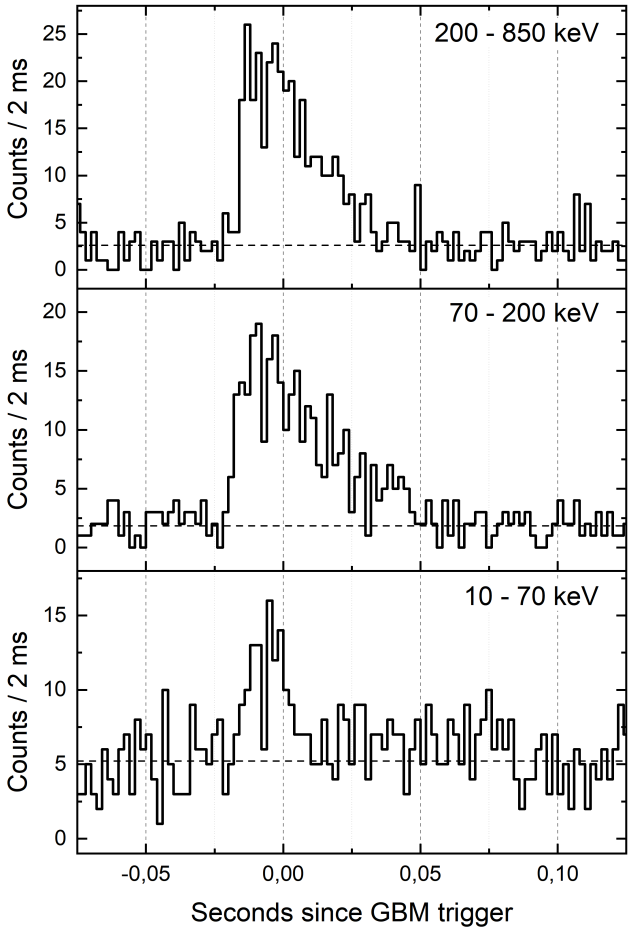}
\caption{Light curves of GRB 231115A based on the {\sl
    Fermi}/GBM experiment data with a time resolution of 2 ms:
  in the range 200--850 keV --- according to data from the
  NaI\_06 -- NaI\_09, NaI\_11, and BGO\_01 detectors, in the
  ranges 10--70 keV and 70--200 keV --- according to data from
  the NaI\_06 -- NaI\_09, and NaI\_11 detectors. The horizontal
  axis shows time in seconds relative to the {\sl Fermi}/GBM
  trigger, the vertical axis shows the number of counts in each
  time bin. The background level is indicated by a dashed line.}
  \label{fig:gbmlc}
\end{figure}

\subsection*{Structure of light curve}
\noindent
The light curves were analyzed using event-by-event (TTE) data
from the most illuminated detectors NaI\_06 -- NaI\_09, NaI\_11,
and BGO\_01 of the {\sl Fermi}/GBM experiment. The light curve in
three channels covering the energy range 10--850 keV is
presented in Fig.~\ref{fig:gbmlc}. The duration parameter
$T_{90}$, the time interval during which the detector records
90\% of the total number of counts (see, for example, Koshut et
al. 1996), was $T_{90} = 65 \pm 1$ ms for GRB 231115A, which is
typical for both giant flares of SGRs and short GRBs. 

It is seen from Fig.~\ref{fig:gbmlc} that the shape of the GRB
231115A light curve varies depending on the energy range --- the
shortest burst is in the soft 10--70 keV range, and the longest
one is in the middle one of 70--200 keV. This is unusual for
gamma-ray bursts, which duration usually decreases with
increasing energy in a power-law manner (Fenimore et
al. 1995)\footnote{This contradicts the conclusion drawn
  from the analysis of the light curves obtained by the
  IBIS/ISGRI telescope of the INTEGRAL observatory, that the
  burst time profile in the softest channel becomes wider than
  that in the hard channels. Note, however, that
  the sensitivity of {\sl Fermi}/GBM at such low energies falls
  noticeably while its background count rate rises.}. The light
curve probably consists of two emission episodes: the main one
(time interval from -0.02 s to 0.01 s) and the secondary one
(interval from 0.01 s till 0.05 s), different in the shape of
the energy spectrum (see the next section).

The minimum time scale of variability, defined as the minimum
time interval during which the energy flux from the source
changes by more than 3 standard deviations compared to the
adjacent intervals, is observed in the middle of the main
episode (at $T \simeq -0.006$ s) and is equal to $\sim 2$
ms. However this cannot be an indicator for a SGR flare since
this kind of rapid variability is also seen in many gamma-ray
bursts (Mitrofanov et al. 1990).
\begin{figure}[!t]
\includegraphics[width=1.05\columnwidth]{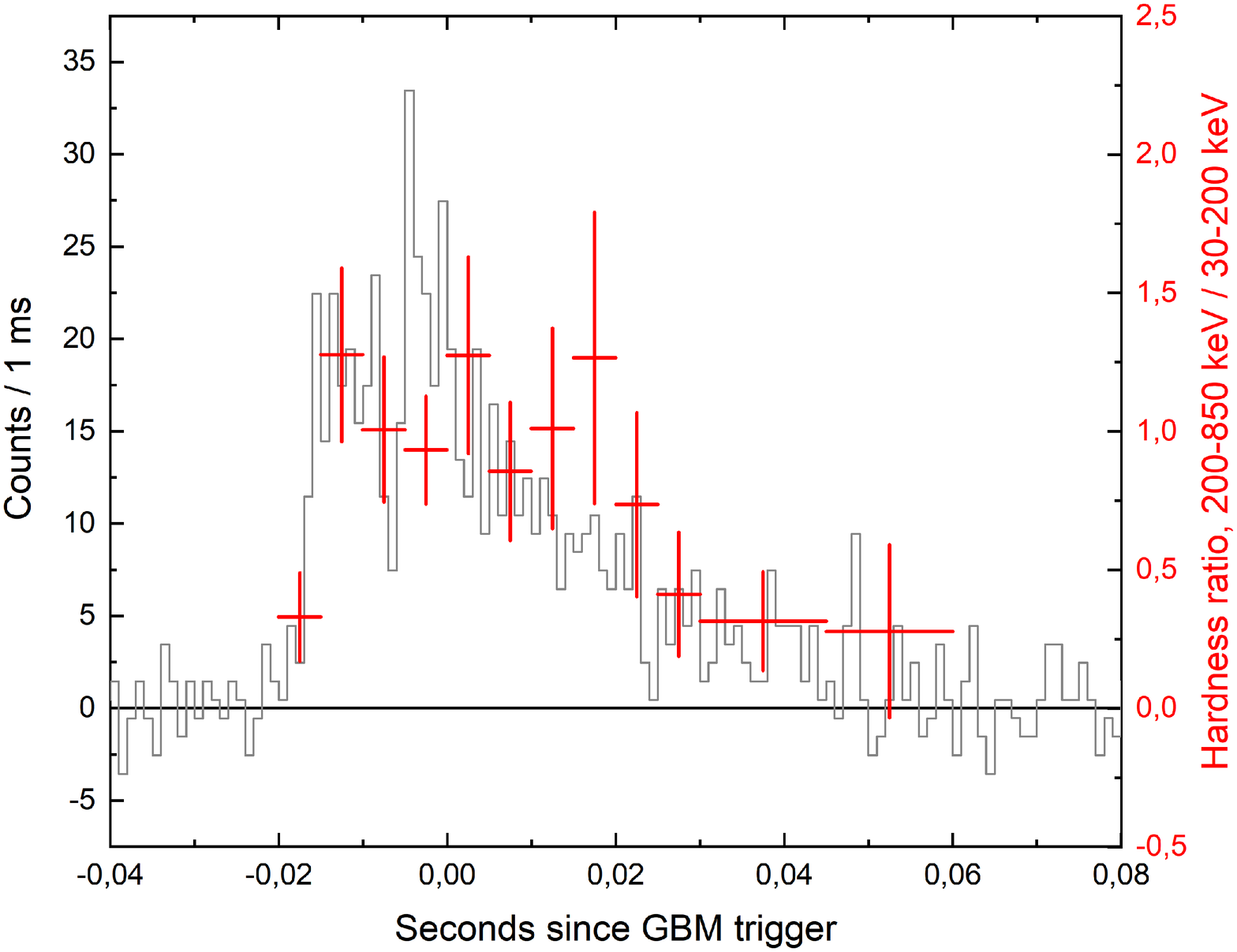}
\caption{Evolution of the hardness ratio of the GRB~231115A
  emission with time according to the {\sl Fermi}/GBM data (red
  dots). The hardness is defined as the ratio of the number of
  counts recorded in the 200--850 keV band to that in the
  30--200 keV band. For comparison, the black line shows the
  light curve of GRB~231115A with a time resolution of 1~ms in
  the 30--850 keV range (according to data from the detectors
  NaI\_06 -- NaI\_09, and NaI\_11, BGO\_01 of {\sl
    Fermi}/GBM). Time in seconds relative to the moment of the
  {\sl Fermi}/GBM trigger is on the horizontal axis, the number
  of counts in the light curve bin is on the vertical left axis
  and the value of the hardness ratio is on the vertical right
  axis.} \label{fig:gbmhr}
\end{figure}
Galactic giant flares of SGRs are also characterized by long (up
to several hundred seconds) extended emission with a detectable
periodicity. Relative contribution of the extended emission to
the total energy of the phenomena varies widely from 1 to 30\%
(Mazetz et al. 2008). We have not detected any significant
extended emission for the event GRB~231115A in the {\sl
  Fermi}/GBM data in both the wide 10--850 keV energy range and
the narrower energy bands 10--70, 70--200, and 200--850 keV.
The upper limit (at the level of three standard deviations) on
the time-integrated flux of extended emission on a time scale of
50 s in the 10--850 keV range exceeds the flux of the main
emission episode of GRB~231115A by almost three times. Thus,
the lack of detection of any extended emission cannot serve as a
reason to reject the association of GRB 231115A with a giant
flare of SGR sources.

\subsection*{Spectral evolution}
\noindent
Gamma-ray bursts are characterized by spectral evolution,
which can be measured by a relative shift (lag) of the light
curve profiles in different energy bands. The lag is considered
positive if the hard emission ``leads'' the soft one, and it is
determined either using cross-correlation analysis of the light
curves (Minaev et al. 2014) or as a shift in the position of the
light curve maximum (Hakkila \& Preece 2011).  The elementary
structures (pulses) in the light curve of gamma-ray bursts are
usually characterized by the positive lag, while the negative
lag observed in some cases may be a consequence of the
superposition effect and arises when the bursts with a complex,
multi-pulse structure of the light curve are analyzed, because
individual pulses have unique properties (Minaev et al.  2014).

In this work, to study the spectral evolution, we used the
cross-correlation method described by Minaev et al. (2014). For
this purpose the light curves of GRB~231115A were constructed
with a time resolution of 1~ms in seven different energy bands
covering the energy range 10--1000 keV. The light curves based
on data from the NaI\_06 -- NaI\_09, and NaI\_11 detectors were
summed, but the light curve constructed from data of the BGO\_01
detector was studied separately.  The 120--220 keV light curve
formed using data from the detectors NaI\_06 -- NaI\_09, and
NaI\_11 was selected as the reference one, that is the curve
relative to which the remaining curves have been
cross-correlated.

The cross-correlation analysis have not revealed significant
spectral evolution, in contrast to GRB~200415A, also apparently
a magnetar giant flare (Minaev \& Pozanenko 2020b).  At the
first glance, the results obtained contradict the behavior of
the light curves in Fig.~\ref{fig:gbmlc}, from which it is
obvious that the energy spectrum evolves with time --- the
profile of the light curve in the soft 10--70 keV band is
noticeably narrower and ends earlier that the pulses in the
middle (70--200 keV) and hard (200--850 keV) bands. This implies
a negative spectral lag. However, the cross-correlation analysis
reveals mainly a shift in the light curves in the region of
their maximum, which is absent in our case. The absence of a
significant spectral lag is probably due to the complex
structure of the light curve and the superposition effect
(Minaev et al. 2014).

Also, we have examined the time evolution of the hardness ratio
of the burst emission using the {\sl Fermi}/GBM data. We have
defined the hardness as the ratio of fluxes expressed in
instrumental counts with subtracted background in the 200--850
and 30--200 keV energy bands. Figure~\ref{fig:gbmhr} shows the
corresponding dependence of the hardness ratio on time. The
dependence generally follows the profile of the burst's light
curve, it rises quickly, remains high for some time during
the main bright episode, and then falls. The figure confirms the
evolution of the hardness ratio found in data of the IBIS/ISGRI
telescope of the INTEGRAL observatory
(Fig.\,\ref{fig:ISGRI_Hardness_Ratio}, note that slightly
different energy bands have been used in the hardness
calculations here). This kind of behavior is observed sometimes
in the gamma-ray bursts with a complex light curve structure
(Gupta et al. 2021), therefore it is impossible to unambiguously
classify this event as a magnetar giant flare based on the type
of this dependence only.
\begin{figure}[!t]
\centering
\includegraphics[width=0.99\columnwidth]{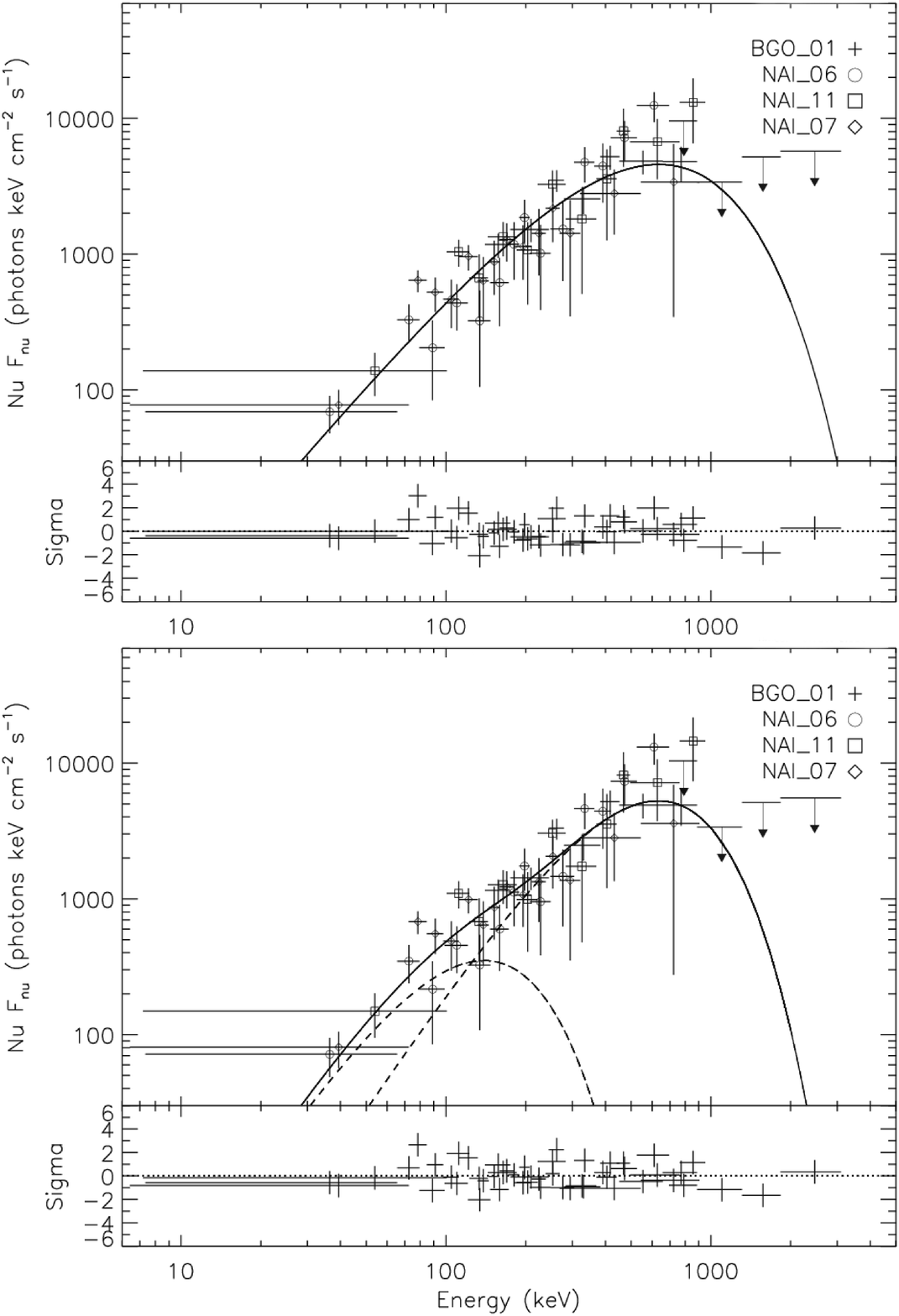}
\caption{Energy spectrum $\nu F_{\nu}$ of GRB~231115A according
  to the {\sl Fermi}/GBM data in the time interval (--0.02 s,
  0.05 s) relative to the trigger time. It is approximated
  either by a power-law model with an exponential cutoff (CPL
  --- the {\sl top\/} figure) or by the sum of two black-body spectra (2\,BB
  --- the {\sl bottom\/} figure). The upper panels in the figures
  show the spectrum obtained from data of the {\sl Fermi}/GBM detectors NaI\_06,
  NaI\_07, NaI\_11, and BGO\_01. The lower panels show 
  deviations of the spectral model from the experimental data in
  standard deviations.}\label{fig:gbmsp}
\end{figure}

\subsection*{Emission spectrum}
\noindent
To reconstruct and fit emission spectra of the burst we used
the RMfit v4.3.2 software package, specially developed for
analyzing data from the GBM monitor of the {\sl Fermi\/}
observatory (fermi.gsfc.nasa.gov/ssc/data/analysis/rmfit/). The
spectral analysis technique is similar to that proposed by
Gruber et al. (2014). The energy spectra were analyzed using
data from the NaI\_06, NaI\_07, NaI\_11, and BGO\_01 detectors
of the {\sl Fermi}/GBM experiment.

We studied the energy spectrum $\nu F_{\nu}$ of GRB~231115A in
three time intervals: interval (--0.02 s, 0.05 s) corresponds to
the spectrum of the whole burst, (--0.02 s, 0.00 s) --- the main
emission episode, (0.00 s, 0.05 s) --- the secondary one. The
energy spectrum in all studied time intervals is
unsatisfactorily described by the thermal black-body model {\sc
  bbody} (BB), the optimal model is a power law with an
exponential cutoff {\sc cutoffpl} (CPL, see equation
(\ref{eq:cpl})).

Although the combination of two thermal models (2\,BB) gives
slightly better approximation of the data, the observed
difference in the value of the statistical functional CSTAT does
not allow one to reject the CPL model that has one degree of
freedom less. The results of our spectral analysis using the BB,
CPL, and 2\,BB models are presented in Table~\ref{tab_spec}.

The emission spectrum $\nu F_{\nu}$ in all cases has the
photon index $\alpha \simeq -0.3$ with the spectral peak energy
$E_{p} \simeq 630$ keV (Table~\ref{tab_spec}). The given value
of the spectral index $\alpha$ is atypical for short gamma-ray
bursts, which on average are characterized by a more rapidly
decaying spectrum with a value of $\alpha \simeq 0.7$ (see, for
example, Burgess et al. 2019). A similar value of the spectral
index was previously observed in other SGR giant flares (Minaev
\& Pozanenko, 2020b). In Fig.~\ref{fig:gbmsp} the spectrum of
the whole burst is presented in the time interval
(--0.02 s, 0.05 s) fitted by the CPL and 2\,BB models.
\begin{table*}[t]
\caption{Results of the spectral analysis of the GRB~231115A
  emission based on {\sl Fermi}/GBM data}
\label{tab_spec}
\vspace{5mm}
\centering
\begin{tabular}{c|c|c|c|c|c|c} \hline\hline
$\Delta$T, \a\ & Model & CSTAT &  $A,$   & $\alpha$\bb\  &
  $E_{p},\ kT_{bb}$, \v\  & Fluence, \g\ \\	
ms 	             &        &  /dof  &   cm$^{-2}$ s$^{-1}$ keV$^{-1}$ &         &  keV     &   $10^{-7}$ erg cm$^{-2}$   \\ \hline
&&&&&&\\ [-3.0mm]
(--20, 50)          & BB    & 485/412 &  (3.33$_{-0.58}^{+0.67}$)$\times$10$^{-6}$    & --        &  135 $\pm$ 8 &  7.50 $\pm$ 0.50 \\
                 &  CPL   & 471/411 & (6.27$_{-0.65}^{+0.72}$)$\times$10$^{-2}$ & -0.34 $\pm$ 0.20 & 637 $_{-58}^{+71}$ & 7.25 $\pm$ 0.46 \\
                & 2BB   &  466/410 &  (4.6$_{-2.1}^{+3.9}$)$\times$10$^{-5}$     &  -- & 36 $_{-7}^{+10}$ & 7.55 $\pm$ 0.50 \\ 
                &    &   &
(1.62$_{-0.46}^{+0.55}$)$\times$10$^{-6}$     &  -- & 162$_{-13}^{+16}$ &  \\ \hline
&&&&&&\\ [-3.0mm]
(--20, 0)          & BB    & 435/412 & (6.3$_{-1.3}^{+1.7}$)$\times$10$^{-6}$        & --       & 132 $\pm$ 10 &  3.71 $\pm$ 0.32 \\
                 &  CPL   & 423/411 &  (1.18$_{-0.15}^{+0.17}$)$\times$10$^{-1}$   &   -0.20 $\pm$ 0.23 & 631 $_{-72}^{+92}$  & 3.50 $\pm$ 0.30 \\
                & 2BB   &  419/410 &  (1.56$_{-0.83}^{+1.80}$)$\times$10$^{-4}$     &  -- & 29 $_{-7}^{+10}$ &  3.72 $\pm$ 0.32 \\
                &    &   &  (3.2$_{-1.0}^{+1.3}$)$\times$10$^{-4}$      &  --  & 156 $_{-15}^{+18}$ &  \\ \hline 
&&&&&&\\ [-3.0mm]
(0, 50)          & BB    & 438/412 & (2.22$_{-0.55}^{+0.71}$)$\times$10$^{-4}$        & --       & 137 $\pm$ 13 &  3.82 $\pm$ 0.38 \\ 
                 &  CPL   & 433/411 & (4.17$_{-0.67}^{+0.78}$)$\times$10$^{-2}$  & -0.47 $\pm$ 0.31 & 635 $_{-84}^{+114}$ &  3.74 $\pm$ 0.35 \\ 
               & 2BB   &  432/410 &  (1.7$_{-1.1}^{+2.7}$)$\times$10$^{-5}$     &  -- & 44 $_{-14}^{+22}$ & 3.85 $\pm$ 0.38 \\ 
               &    &   &   (9.9$_{-4.9}^{+5.9}$)$\times$10$^{-7}$    &  --  & 168 $_{-22}^{+32}$ &  \\ \hline              

\multicolumn{7}{l}{}\\ [-1mm]
\multicolumn{7}{l}{\a\ Time interval relative to trigger of
  {\sl Fermi}/GBM.}\\ 
\multicolumn{7}{l}{\bb\ Photon index of the power-law component.}\\
\multicolumn{7}{l}{\v\ Spectral peak energy
  $E_{p}=E_{c}(2-\alpha)$ where $E_c$ is the energy of an
  exponential cutoff in Eq.~(\ref{eq:cpl})}\\
\multicolumn{7}{l}{ \ \ \   or $kT_{bb}$ for the models BB and 2\,BB.}\\ 
\multicolumn{7}{l}{\g\ Time-integrated energy flux in the
  range 10--1000 keV.}\\ 
\end{tabular}
\end{table*}

Spectral analysis of two episodes of GRB~231115A, identified
during the analysis of light curves, did not reveal significant
differences in the parameters of the spectral models. Within the
CPL model, the photon index $\alpha$ of the secondary episode is
slightly higher with nearly the same value of the exponential
cutoff ($E_{p} \simeq 630$ keV), which may explain the relative
lack of the soft emission in the secondary episode.

We estimated the isotropic equivalent of the total energy
$E_{iso}$ under the assumption of association of the source with
the galaxy M\,82 ($D_L = 3.5$ Mpc) in the energy range
1--10\,000 keV, having extrapolated the CPL model spectrum
obtained in the 7--3000 keV range. It is equal to $E_{iso} =
(1.28\pm 0.14)\times 10^{45}$ erg. Likewise, the
peak luminosity calculated in the time interval (--0.02 s, 0.0 s) is
$L_{iso} = (3.09 \pm 0.44)\times 10^{46}\ \mbox{\rm erg
  s}^{-1}$.

For the thermal black-body model (BB) one can estimate the
radius of the emitting region using the peak luminosity $L_{iso}
= 3.09\times 10^{46}\ \mbox{\rm erg s}^{-1}$ and the
Stefan-Boltzmann law for the emission with the temperature
$kT_{bb} = 132$ keV. The resulting radius $R_{bb}\simeq 27$ km
corresponds by order of magnitude to the size of the
magnetosphere of a neutron star.
\begin{figure*}[t]
\centering
\includegraphics[width=0.8\linewidth]{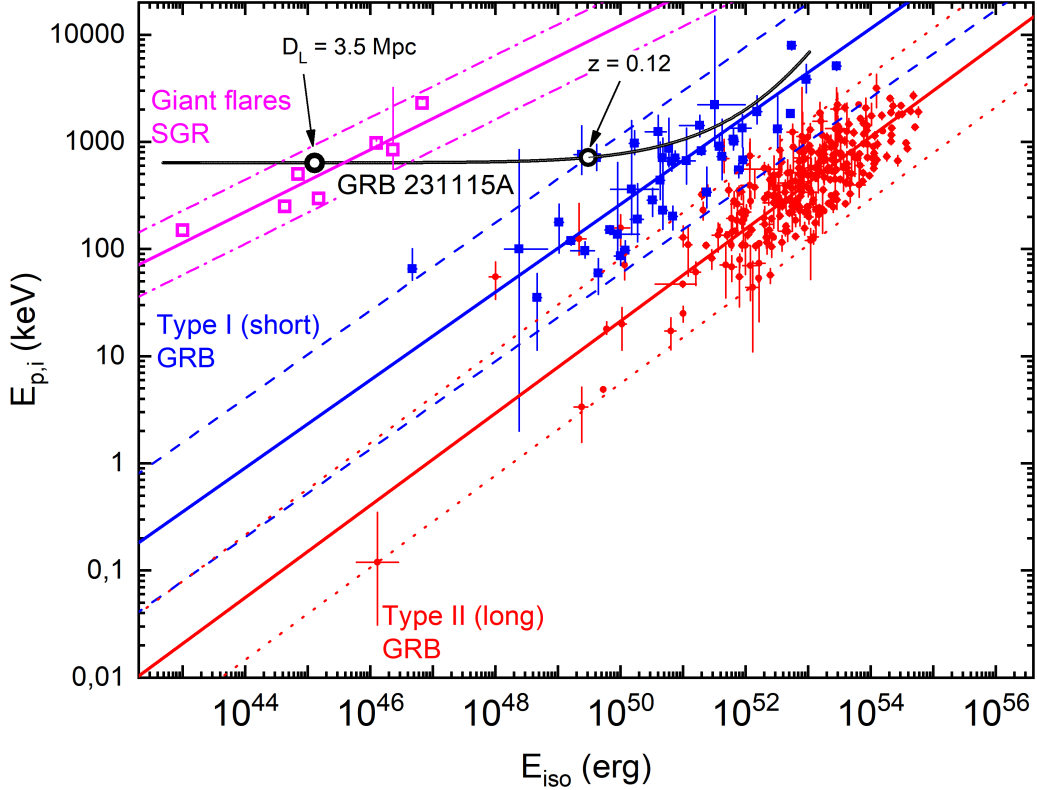}
\caption{The $E_{p,i}$ -- $E_{iso}$ correlation for type I (blue
  squares) and type II (red circles) GRBs and for SGR giant
  flares (magenta unfilled squares) with corresponding fitting
  results, including $2 \sigma_{cor}$ correlation boundaries,
  shown in corresponding colors. The black curve shows the
  trajectory of GRB~231115A as a function of redshift, the
  position for $D_L = 3.5$ Mpc and $z = 0.12$ is shown by
  unfilled black circles.} \label{fig:amati}
\end{figure*}

\section*{POPULATION ANALYSIS OF THE BURST}
\noindent
It is interesting to compare the obtained values of the duration
and energetics of GRB~231115A with the parameters of other
gamma-ray transients.
\subsection*{$E_{p,i}$ -- $E_{iso}$ correlation}
\noindent
Minaev \& Pozanenko (2020a) have shown that the correlation
between the isotropic equivalent of the total energy, emitted in
the gamma-ray range, $E_{iso}$ and the position of the extremum
in the energy spectrum $\nu F_{\nu}$ in the source's rest frame
$E_{p,i}$ (equation (\ref{eq:amati}), Golenetskii et al. 1983;
Amati et al. 2002) can be effectively used to classify GRBs.

This is facilitated by the observational fact that the
correlation for various types of gamma-ray bursts is described
by a power law with a single index of $a \simeq 0.4$, while the
correlation region of type I GRBs is located upper than the
correlation region of type II bursts. Subsequently, in Minaev \&
Pozanenko (2020b) the known giant flares of SGR sources were
additionally considered, and it turned out that they also follow a
similar correlation with the index $a \simeq 0.3$ and occupy an
isolated location on the $E_{p,i}$ -- $E_{iso}$ diagram. This
made it possible to include them in the classification scheme
proposed by Minaev \& Pozanenko (2020a).
\begin{equation}\label{eq:amati}
    \lg\Big(\frac{E_{p,i}}{100~\mbox{\rm keV}}\Big) =
    a\lg\Big(\frac{E_{iso}}{10^{51}~\mbox{\rm erg}}\Big) + b.
\end{equation}
To analyze the position of GRB~231115A on the $E_{p,i}$ --
$E_{iso}$ diagram, we used a sample of 316 gamma-ray bursts and
7 magnetar giant flares, as well as the results of analysis
of the $E_{p,i}$ -- $E_{iso}$ correlation for this sample
(Minaev \& Pozanenko 2020a,b, 2021). The corresponding $E_{p,i}$
-- $E_{iso}$ diagram is presented in Fig.~\ref{fig:amati}. It is
obvious that the location of GRB~231115A on the diagram allows
us to unambiguously classify the burst as a magnetar giant
flare. The figure also shows the trajectory of changing the
source location on the diagram vs its redshift $z$. The event
could be also classified as a short gamma-ray burst since
redshift $z = 0.12$ (the intersection point of the trajectory
with the boundary of the short gamma-ray burst correlation
region).

\subsection*{$T_{90,i}$ -- $EH$ diagram}
\begin{figure*}[t]
  \centering
\includegraphics[width=0.8\linewidth]{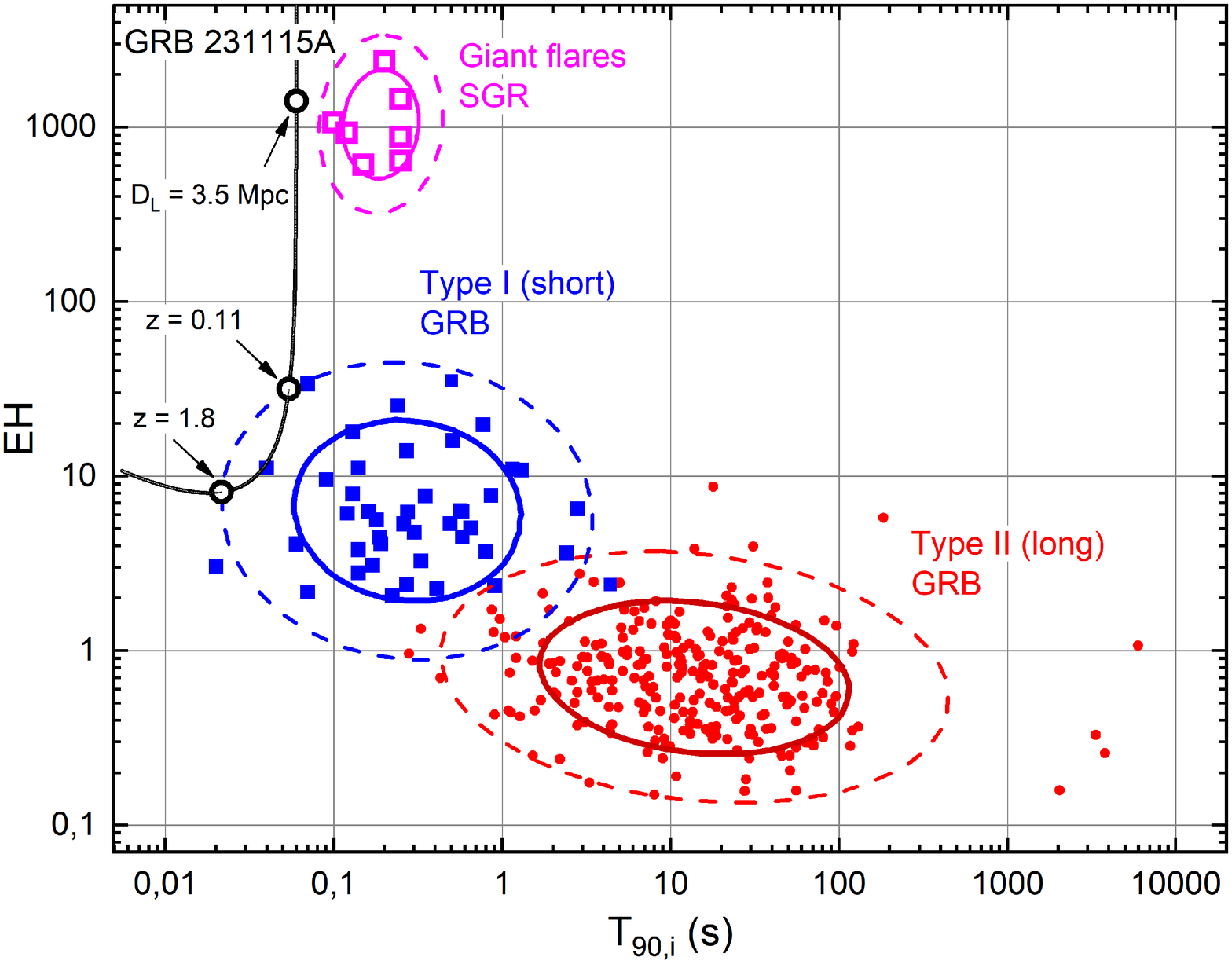}
\caption{The $T_{90,i}$ -- $EH$ diagram for type I (blue
  squares), type II (red circles) GRBs and SGR giant flares
  (magenta unfilled squares) with corresponding results of the
  cluster analysis, the $1\,\sigma_{cor}$ and $2\,\sigma_{cor}$
  regions of the clusters are shown by thick solid and thin
  dashed curves of the corresponding colors. The black curve
  shows the trajectory of GRB~231115A as a function of
  redshift. The position for $D_L = 3.5$ Mpc, as well as the
  positions on the trajectory, consistent with the suggestion
  that the event was a short gamma-ray burst, at $z = 0.11$ and
  $z = 1.8$ are marked with unfilled black circles.} \label{fig:ehd}
\end{figure*}
\noindent
To solve the problem of gamma-ray burst classification, another
method was proposed by Minaev \& Pozanenko (2020a), which, in
addition to features of the $E_{p,i}$ -- $E_{iso}$ correlation,
uses the bimodality of the duration distribution of GRBs in the
rest frame $T_{90,i}$. For this purpose, the $EH$ parameter was
introduced (equation (\ref{eq:EH})), characterizing the location
of the gamma-ray burst on the $E_{p,i}$ -- $E_{iso}$ diagram:
\begin{equation}\label{eq:EH}
    EH = \frac{(E_{p,i} / 100~\mbox{\rm keV})}{ (E_{iso} /
      10^{51}~\mbox{\rm erg})^{~0.4}}.	
\end{equation}

Figure~\ref{fig:ehd} shows the $T_{90,i}$ -- $EH$ diagram for
316 gamma-ray bursts and 7 giant flares of magnetars from
(Minaev \& Pozanenko 2020a,b, 2021). Type I GRBs compared to
type II ones have a higher spectral hardness value $E_{p,i}$ at
a lower value of the total energy $E_{iso}$ and, as a
consequence, a larger value of the parameter $EH$, and shorter
duration $T_{90,i}$. SGR giant flares have the same duration as
type I GRBs but much lower energetics with the similar spectral
hardness which results in a large value of the $EH$
parameter. Thus, the $T_{90,i}$ -- $EH$ diagram can also be used
not only to classify GRBs, but also to separate giant flares of
SGRs from type I GRBs. GRB 231115A is clearly classified as a
giant flare of SGR, being in close proximity to the
corresponding cluster of events (parameter $EH = 1450$).  The
figure also shows the trajectory of the event on the diagram as
a function of its source redshift. Note that the event can be
classified as a short gamma-ray burst in the redshift range from
$z = 0.11$ to $z = 1.8$ (points of the intersection of the
trajectory with boundaries of the short GRB cluster region).

\section*{OPTICAL OBSERVATIONS}
\noindent
Among the optical observation networks, the first to respond to
the notification on registration of GRB\,231115A from {\sl
  Fermi}/GBM (Dalessi et al. 2023) was the GROWTH-India survey,
which began observing the localization area since 2023-11-15
16\uh47\um58\fs140 UT. Kumar et al. (2023a) reported the
detection of the candidate AT\,2023xvj in the optical afterglow
of GRB\,231115A in the wing of the M\,82 galaxy 1.19 hours after
the {\sl Fermi}/GBM trigger. At the time of discovery, the
source had coordinates R.A.$ = 09\uh56\um00\fs2 \pm 0\farcs6$,
Decl.$ = 69\deg40\arcmin29\farcs2 \pm 0\farcs6$ (J2000 epoch)
and brightness $\sim 19.2$ mag in the photometric band
$r^{\prime}$.  Apart of GROWTH-India, optical observations to
search for the optical component of GRB~231115A have been
carried out by many other instruments (see section
Observations). However, the source announced in the circular by
Kumar et al. (2023a) has not been found. Kumar et al. (2023b)
later reported on an error in their previous analysis published
in Kumar et al. (2023a). The error occurred when subtracting the
host galaxy template image from the PanSTARRS DR1 survey
(Chambers et al. 2016) which contains artifacts.
\begin{table*}[t]
\centering
\caption{GRB-IKI-FuN telescopes used in observations of GRB 231115A}
\label{tab:optical_telescopes}

\vspace{4mm}
\begin{tabular}{c|c|c|r@{$\times$}l|l}\hline\hline
 Observatory& Telescope& D, m \a&\multicolumn{2}{c|}{FoV\bb}&Location\\
 \hline
 &&&\multicolumn{2}{c|}{}&\\ [-3.0mm]
       AbAO/GENAO&  AS-32&  0.7&  44\farcm4&44\farcm4 &Abastumani, Georgia\\
ISON-Kitab& RC-36& 0.36&43\farcm7&43\farcm7&Kitab, Uzbekistan\\ \hline
\multicolumn{6}{l}{}\\
\multicolumn{6}{l}{\a\ Diameter of the telescope mirror.}\\
\multicolumn{6}{l}{\bb\ Field of view of the telescope.}\\     
\end{tabular}
\end{table*}
\begin{table*}[t]
\centering
\caption{Log of optical observations of the gamma-ray burst GRB~231115A}
\label{tab:opt_obs_log}
\vspace{4mm}
\begin{tabular}{c|c|c|c|c|c|c|c}
\hline\hline
$T-T_0,$\a & $\Delta T,$\bb & Upper limit,\v & $f_{\nu}$,\g & Filter &  Observatory & Telescope & GCN \\
days & s & mag &  $\mu$Jy  & & & & \\ \hline
&&&&&&&\\ [-3.0mm]
-1.155810 & --- & 20.47 & 23.6 & r & ZTF & Palomar 1.2m & 35048 \\
-0.452917 & $12\times60$ & 20.1 & 27.9 & ~L\dd\ & ORM/SSO & GOTO & 35050 \\
0.054688 & 180 & 18.8 & 92.5 & Clear\e\ & MASTER-Tunka & 0.4m & 35046 \\
0.075023 & 600 & 19.5 & 48.6 & Rc & MITSuME & Akeno 0.5m & 35057 \\
0.087118 & 1200 & 19.8 & 36.8 & Rc & MITSuME & Akeno 0.5m & 35057 \\
0.111250 & 2400 & 20.2 & 25.5 & Rc & MITSuME & Akeno 0.5m & 35057 \\
0.144980 & 7380 & 19.9 & 33.6 & Rc & MITSuME & Okayama 0.5m & 35057 \\
0.149271 & $19\times30+123\times60$ & ~~18.6\,\zg & 111 & Clear\e &
ISON-Kitab & RC-36 & This work \\
0.418507 & 7200 & 22.0 & 5.75 & r & WO & Fraunhofer 2m & 35092 \\
0.500000 & --- & 22.0 & 5.75 & r & ORM & TNG & 35077 \\
0.447095 & $89\times60$ & ~~19.3\,\zg & 58.4 & R & AbAO & AS-32 &
This work\\
0.500000 & $2\times195$ & 21.6 & 8.32 & r & ORM & Liverpool 2m & 35067 \\
\hline
\multicolumn{8}{l}{}\\
\multicolumn{8}{l}{\a\ Exposure start time relative to the
  {\sl Fermi}/GBM trigger.}\\
\multicolumn{8}{l}{\bb\ Exposure duration.}\\
\multicolumn{8}{l}{\v\ The upper limit at the level of 3 standard deviations.}\\
\multicolumn{8}{l}{\g\ The upper limit on the flux density at
  the level of 3 standard deviations.}\\ 
\multicolumn{8}{l}{\dd\ GOTO wide-band L filter (4000--7000 \AA).}\\
\multicolumn{8}{l}{\e\ Clear light (unfiltered).}\\
\multicolumn{8}{l}{\zg\ Not corrected for extinction in the Galaxy of $E(B-V)=0.1326$ (Schlafly \& Finkbeiner 2011).}\\
\end{tabular}
\end{table*}

Two scientific groups have discovered several candidates in
transient sources in the localization region of the burst by the
IBAS system of the INTEGRAL observatory. Thus, Perley et
al. (2023) discovered a red source with coordinates
R.A. $=09\uh55\um53\fs07$, Decl. $=+69\deg40\arcmin23\farcs0$,
and Hu et al. (2023) revealed another red source W231115b with
coordinates
R.A. $=09\uh55\um58\fs81$,\ Decl. $=69\deg41\arcmin28\farcs5$,
which had brightness $r = 21.26 \pm 0.08$. Hu et al. (2023) also
discovered a source subsequently found by Perley et al. (2023),
which was designated as W231115a. The apparent magnitude of this
source was $r = 20.54\pm 0.04$. Apparently these sources are in
the disk of the M\,82 galaxy, but not associated with
GRB~231115A. Several reasons can be given why these sources were
not discovered earlier: a significant gradient of background
emission from the M\,82 galaxy, a high density of sources in the
galactic disk, significant absorption of optical emission by
dust, and insufficient sensitivity of observing instruments.

In turn, we also carried out observations using the GRB-IKI-FuN
({\sl IKI Gamma-Ray Burst Follow-up Network\/}, Volnova et
al. 2021) telescope network in order to search for a possible
optical component of GRB~231115A. The list of telescopes used
and their main parameters are presented in
Table~\ref{tab:optical_telescopes}.

Observations with the AS-32/AbAO telescope were taken in the
photometric R band, while observations with the RC-36/ISON-Kitab
were carried out in the white light (without
filters)\footnote{{\it Added in proofs:\/} After passing the
  manuscript in press, observations of the M\,82 galaxy were
  made in its quiet state (when a possible optical component
  should have already faded) on December 8 and 9, 2023
  ($\sim21.5$ days after the burst) on the telescopes in AbAO
  and Kitab, respectively. This allowed us to more accurately
  subtract the contribution of the M\.82 galaxy from the images
  we obtained immediately after the burst.}. The log of optical
observations is presented in Table~\ref{tab:opt_obs_log}.

Observational data (images) were processed in a homogeneous
manner using units of the pipeline for optical transient
search APEX v2023.11 (Pankov et al. 2022). According to the
methodology described by Pankov et al. (2022), the images
underwent quality control, calibration, stacking, astrometry and
differential photometry. The search for variable sources was
carried out by subtracting the template image of the host galaxy
M\,82, taken with the same instrument, but at a later epoch. For
this purpose, a special pipeline unit APEX
\texttt{apex\_subtract} was developed. Due to the large angular
size of M\,82 in astronomical images (of about
$4\arcmin\times4\arcmin$), template subtraction was a necessary
procedure to remove the significant background gradient from the
M\,82 galaxy.

\begin{figure*}[!t]
    \centering
    \includegraphics[width=0.8\linewidth]{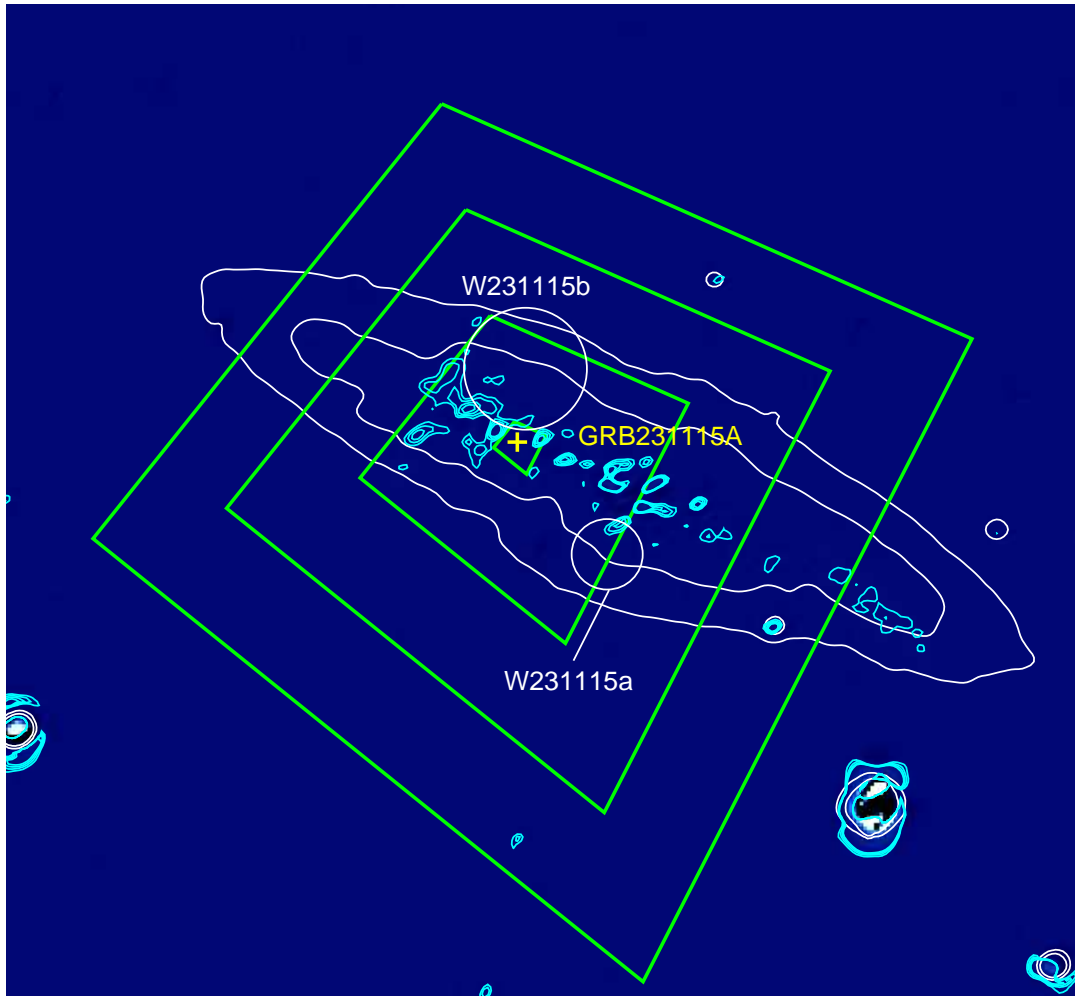}
    \caption{The same image of the GRB\,231115A localization
      region as in Fig.~\ref{fig:ibis-smallmap}, but after
      subtracting the contribution of the M\,82 galaxy in a
      quiet state (observed with the AS-32 telescope on December
      8, 2023, when the possible optical component of the burst
      should have already faded). As before, the green diamonds
      show the contours of the localization of the burst in the
      X-ray range, the third one from the center corresponds to
      the 90\% confidence level ($1\farcm5$ uncertainty). White
      circles are two proposed optical candidates (Hu et
      al. 2023, see text). White thin contours indicate the
      profile of the M\,81 galaxy. The blue contours are
      residuals at the place of the galaxy after subtraction
      ({\sl added in proofs\/}).} \label{fig:as32-smallmap}
\end{figure*}

After subtraction, we did not find reliable candidates to the
optical afterglow of GRB 231115A in the localization region
obtained in the IBIS/ISGRI INTEGRAL experiment
(Fig.~\ref{fig:as32-smallmap}). Also, the sources from the GCN
circulars by Perley et al. (2023) and Hu et al. (2023) were not
detected due to the insufficient sensitivity reached in our
observations.

\begin{figure*}[t]
\centering
\includegraphics[width=0.9\linewidth]{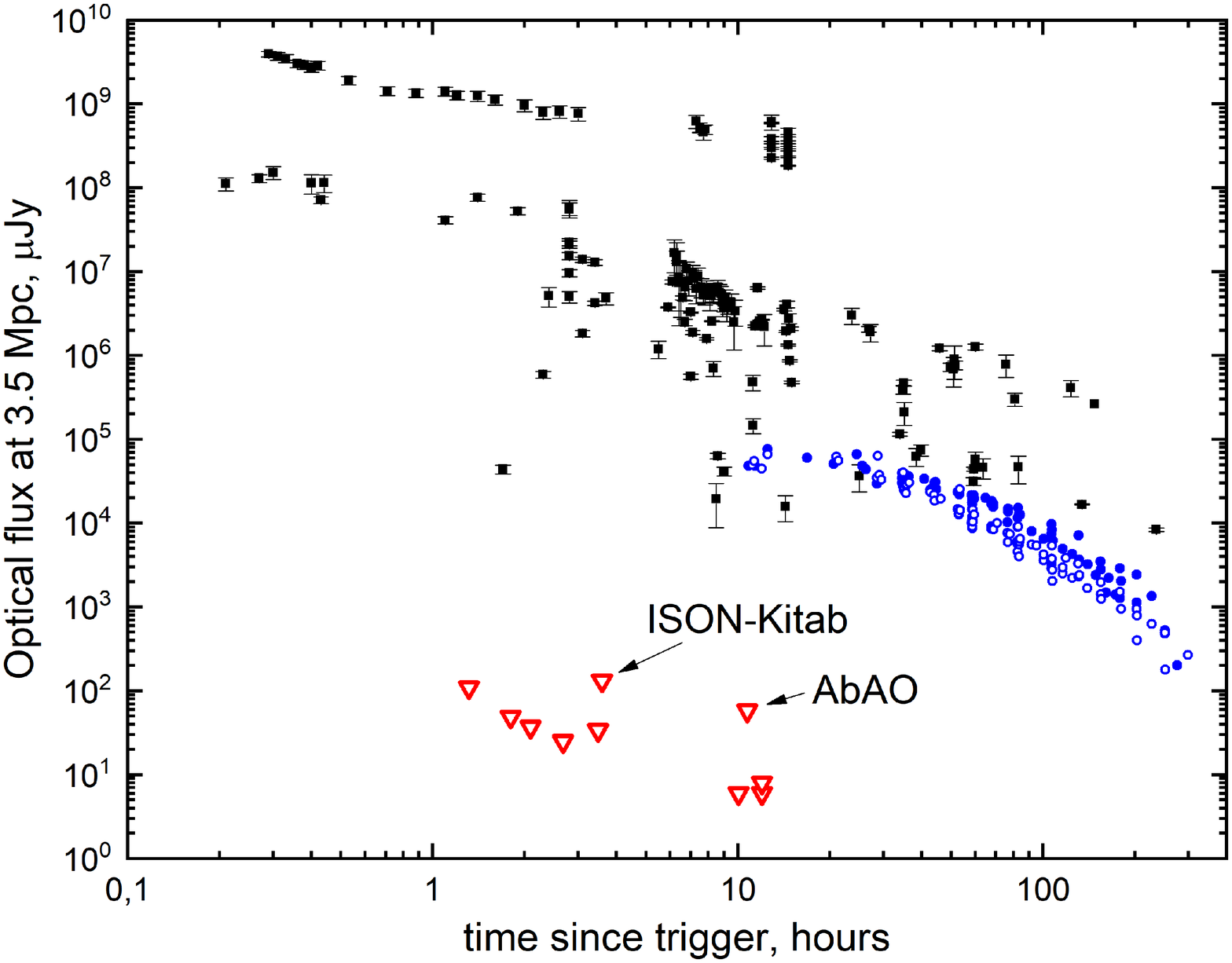}
\caption{Optical observations of GRB~231115A in comparison with
  light curves of other short gamma-ray bursts reduced to a
  luminosity distance of $D_L = 3.5$ Mpc. The horizontal axis
  shows time in hours relative to the {\sl Fermi}/GBM trigger,
  the vertical axis shows the observed flux in $\mu$Jy. The red
  unfilled triangles show upper limits on the optical flux from
  GRB~231115A according to the data from Table~\ref{tab:opt_obs_log},
  the black squares --- results of optical observations of short
  GRBs with measured redshifts from Fong et al. (2015), the blue
  filled and unfilled circles --- the light curve of GRB~170817A
  in filters $r$ and $i$, respectively, according to Villar et
  al. (2017).}  \label{fig:optlim}
\end{figure*}

In Fig.~\ref{fig:optlim} the red triangles show the upper limits
on the possible optical component of GRB 231115A, obtained both
in our observations (marked with arrows) and by other scientific
groups (Table~\ref{tab:opt_obs_log}). In addition,
Fig.~\ref{fig:optlim} presents the results of optical
observations of other short gamma-ray bursts with a known
redshift according to Fong et al. (2015) and the light curve of
the kilonova GRB~170817A according to Villar et al. (2017),
reduced to the luminosity distance $D_L = 3.5$ Mpc.

The deepest upper limits on the optical afterglow of GRB~231115A
were obtained in 10 hours after the trigger ($\simeq 22$ mag in
the $r$ filter). The optical afterglows of other short gamma-ray
bursts (including the optical component of the kilonova GRB
170817A), observed from a distance of $D_L = 3.5$ Mpc, would be
brighter at this time moment by no less than 2500 times (their
brightness would be 13.5 mag). Thus, the obtained upper limits
on the optical component reliably exclude the interpretation of
the GRB 231115A event as a short gamma-ray burst. Note however
that the optical emission could be affected by very strong
absorption along the line of sight if the source of the burst is
located in the farthest (from the observer) part of the disk of
the host galaxy. Unfortunately, the source localization accuracy
is not sufficient to rule out this possibility.

\section*{CONLCUSIONS}
\noindent
A detailed analysis of the spectral and temporal properties of
the short burst GRB~231115A was performed in the hard X-ray and
gamma-ray range using data from the INTEGRAL and {\sl Fermi\/}
space observatories in order to establish the nature of this
burst. Early optical observations were also carried out by
telescopes of the GRB-IKI-FUN network in an attempt to find an
afterglow of the burst or an optical source associated with the
burst.

In particular, the following results have been obtained:

1). The early localization of the burst (performed within the
framework of the Quick Look analysis by the automatic IBAS
system) was confirmed and refined 
(R.A.$=09\uh55\um59\fs28,$\ Decl.$=+69\deg41\arcmin02\farcs40$,
epoch 2000.0, the uncertainty is smaller than $1\farcm5$) as well as its
association with the nearby Cigar galaxy (M\,82) located at a
distance of $D_L = 3.5$ Mpc.  This allows us to consider the
hypothesis of the magnetar origin of the burst, i.e. that a
giant flare of a previously unknown soft gamma repeater (SGR)
occurred in this galaxy on November 15, 2023, as highly
probable. If the hypothesis is confirmed, this would be the
first giant flare of an extragalactic SGR that is well localized
and reliably identified with a known galaxy.

2). The hard X-ray and gamma-ray light curves of GRB~231115A
have the traditional FRED profile (fast rise --- exponential
decay) in all instruments and in all energy bands, with the
exception of the softest (20--50 keV) band of the IBIS/ISGRI
telescope of the INTEGRAL observatory, in which the burst
profile was wider and had a flat top with $\Delta T\simeq60$
ms. Long-lasting (tens of seconds) extended emission typical for
magnetar giant flares has not been detected for GRB 231115A. At
the same time, the obtained upper limit on the time-integrated
flux of the extended emission does not exclude the association
of the burst with a magnetar giant flare.

3). Cross-correlation analysis of the light curves in different
energy bands according to the {\sl Fermi}/GBM monitor data did
not reveal any significant spectral lag. Such behavior occurs in
both short gamma-ray bursts and magnetar giant flares. At the
same time, a noticeable evolution of the hardness ratio was
discovered (in both the {\sl Fermi}/GBM data and IBIS/ISGRI data
of the INTEGRAL observatory). The hardness ratio was 2--3 times
higher (and even increased according to IBIS/ISGRI) during the
first $\sim 40$ ms of the burst, and then steadily decreased
(during the next $\sim80$--$100$ ms).
\begin{figure*}[!t]
\centering
\includegraphics[width=0.84\linewidth]{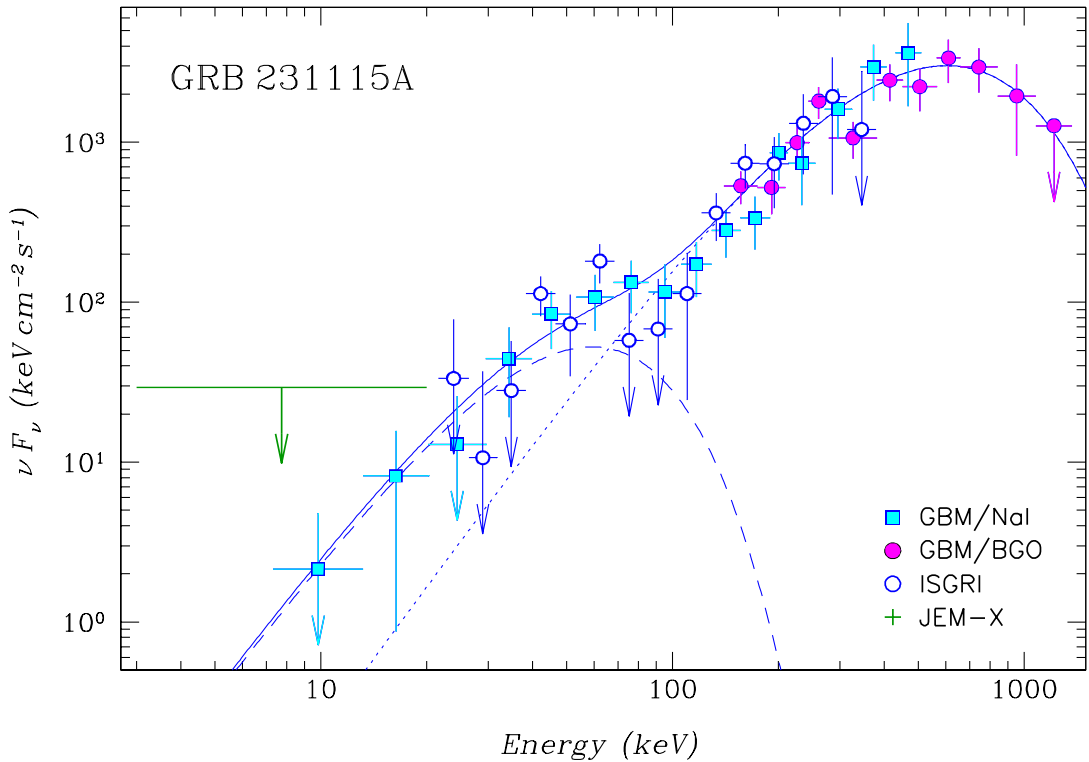}
\caption{Broadband X-ray and gamma-ray spectrum ($\nu F_{\nu}$)
  of GRB\,231115A based on data from four instruments:
  IBIS/ISGRI and JEM-X of the INTEGRAL observatory and GBM (its
  BGO\_01 and NaI\_06 detectors) of the {\sl Fermi\/}
  observatory (the 3--1500 keV energy range). The JEM-X upper
  limit corresponds to one standard deviation. The spectrum was
  accumulated during the total duration of the burst
  (120~ms). The solid line represents the result of the best
  approximation of the spectrum by the {\sc BBR$+$CPL} model,
  the dashed and dotted lines represent individual components of
  the model.}
    \label{fig:all-spec}
\end{figure*}

4). The emission spectrum of the burst contained two
components: a hard one, which was satisfactory described by a
power-law model with an exponential cutoff at high energies
(CPL) typical for gamma-ray bursts and magnetars, and the soft
one, which was approximated with the spectrum of a black-body
photosphere with a temperature of $kT_{bb}\sim 15$
keV. Figure~\ref{fig:all-spec} shows the broadband (3--1500 keV)
emission spectrum of GRB\,231115A, reconstructed from
observational data of the IBIS/ISGRI and JEM-X telescopes of the
INTEGRAL observatory and the gamma-ray burst monitor GBM (its
BGO\_01 and NaI\_06 detectors) of the {\sl Fermi\/}
observatory. Table\,\ref{tab:all-spec} presents results of the
spectral approximation by two models: 2\,BBL and BBL$+$CPL (the
latter model is shown in the figure by a solid curve, and its
components are dashed and dotted lines). It follows from the
figure that the model successfully describes the measured
spectrum of the burst.

The representation of the burst spectrum in the form of $\nu\,
F_\nu(\nu)$ clearly shows that the main energy of its emission
is contained in the photons with $h\nu\sim 500-600$ keV. It is
important that the photon index of the hard emission component
had a peculiar value of $\alpha \simeq -1.06$, ensuring a rapid
increase of the spectral density $F_{\nu}\sim E^{2.06}$ with
photon energy. It is due to this almost Rayleigh-Jeans spectral
index the hard emission of this burst was successfully described
by a black-body spectrum. Such hard power-law ``tails'' in the
emission spectra are more typical for giant flares of
magnetars than for short gamma-ray bursts.
\begin{table*}[t]
\centering
\caption{Results of our analysis of the broadband X- and
  gamma-ray spectrum of GRB 231115A in the 3--1500 keV range
  according to data from the INTEGRAL and {\sl Fermi\/}
  observatories}
\label{tab:all-spec}
\vspace{5mm}
\begin{tabular}{c|c|r@{$\pm$}l|r@{$\pm$}l|r@{$\pm$}l|c} \hline\hline
Model & $\chi^2/N$\,\a\ &\multicolumn{2}{c|}{A}&\multicolumn{2}{c|}{$\alpha$\bb}&
\multicolumn{2}{c|}{$kT_{bb},\ E_{c},$\dd} &Flux\e, \\	
       &&\multicolumn{2}{c|}{}&\multicolumn{2}{c|}{}&\multicolumn{2}{c|}{keV}
&$10^{-6}$ erg s$^{-1}$ cm$^{-2}$  \\ \hline
&&\multicolumn{2}{c|}{}&\multicolumn{2}{c|}{}&\multicolumn{2}{c|}{}&\\ [-3.0mm]
BBR&112.4/140&$13.0$&$0.5$\g\    &\multicolumn{2}{c|}{--}&$135$&$4$&$4.68\pm0.48$\\ [-1.3mm]
$+$&&\multicolumn{2}{c|}{}&\multicolumn{2}{c|}{}&\multicolumn{2}{c|}{}&\\ [-1.3mm]
BBR&&$152$&$27$\g\  &\multicolumn{2}{c|}{--}&$16.0$&$1.5$& $0.13\pm0.04$\\ \hline
&&\multicolumn{2}{c|}{}&\multicolumn{2}{c|}{}&\multicolumn{2}{c|}{}&\\ [-3.0mm]
CPL&113.2/139&$2.58$&$0.67$\v\ &$-1.08$&$0.02$         &$196$ &$8$ &$6.21\pm0.50$\\ [-1.3mm]
$+$&&\multicolumn{2}{c|}{}&\multicolumn{2}{c|}{}&\multicolumn{2}{c|}{}&\\ [-1.3mm]
BBR&       &$160$&$29$\g\    &\multicolumn{2}{c|}{--}&$15.1$&$1.5$&$0.11\pm0.04$\\ \hline
\multicolumn{9}{c}{}\\ [-1mm]
\multicolumn{9}{l}{\a\ Minimum value of $\chi^2$ and the number of degrees of freedom $N$.}\\
\multicolumn{9}{l}{\bb\ Photon index of the power-law component $I_{100}\ (E/100\ \mbox{\rm keV})^{-\alpha}$.}\\
\multicolumn{9}{l}{\v\ Normalization of this component
  $I_{100}$ at 100 keV [$10^{-2}\ \mbox{\rm phot s}^{-1}$ cm$^{-2}$ keV$^{-1}$].}\\
\multicolumn{9}{l}{\g\ Radius of the photosphere $R_{bb}$
  [km] for a distance of $d_{L}=3.5$ Mpc.}\\
\multicolumn{9}{l}{\dd\ Temperature $kT_{bb}$ or the energy of
  an exponential cutoff $E_{c}$.}\\
\multicolumn{9}{l}{\e\ Energy flux in the range of 10--1000 keV.}\\
\end{tabular}
\end{table*}

5). A combination of black-body spectra that allows one to
approximate well the emission spectrum of GRB 231115A, may
arise in hypothetical models of evaporation of primordial
black holes (e.g., Fegan et al. 1978) or the fall of a
primordial black hole of a limited mass onto a supermassive black
hole (Barco et al. 2021). These models can generate very short
hard gamma-ray bursts apparently without afterglow in the X-ray
and optical ranges. However, the total energy emitted in the
gamma-ray range in these models turns out to be significantly
less than the energy measured from GRB~231115A, which would not
allow observing such a flare in the M\,82 galaxy.

6). The obtained upper limits on the flux of the burst's optical
afterglow 3.6 and 10.7 hours after this event are several orders
of magnitude below the level that would be expected from the
short gamma-ray bursts associated with neutron star mergers
(Kann et al. 2011; Villar et al. 2017; Pandey et al. 2019). Our
upper limits are consistent with optical observations with other
telescopes, as well as the lack of any afterglow in the soft
X-ray range. All this confirms the magnetar version of the
origin of GRB~231115A.

7). The position of GRB~231115A on the $E_{p,i} - E_{iso}$ and
$T_{90,i} - EH$ diagrams confirms the classification of the
burst as a SGR giant flare. With acceptable for type I GRBs
duration $T_{90,i} = 0.06$ s and spectral peak energy
$E_{p,i}\sim 640$ keV, the isotropic total energy $E_{iso}\sim
10^{45}$ erg emitted during the burst in the gamma-ray range is
however typical only for SGR giant flares.

8). The distance to the host galaxy of the burst suggests a
confident detection of the gravitational wave signal from the
merger of neutron stars (the cause of short gamma-ray bursts),
which, however, has not been detected by the LIGO/Virgo/KAGRA
detectors. This is the strongest argument in favor of the
magnetar origin of the burst and against the possibility of its
explanation by the merger of a pair of neutron stars that
occurred in this galaxy. There is a very small probability of a
random coincidence of the location of a short distant gamma-ray
burst with the location of this galaxy. This probability can be
estimated as the ratio of the apparent area $11\farcm2\times
4\farcm3$ of the M\,82 galaxy to the total area of the sky
$p\simeq (11\arcmin\times
4\arcmin)/(4\pi)=11\arcmin\times4\arcmin\ (\pi/180/60\arcmin)^2/(4\pi)\simeq
3\times10^{-7}$ or 3\,400\,000 to 1 in favor of the magnetar
origin of the burst. A more conservative estimate of the
probability in favor of the magnetar hypothesis was given by
Burns (2023), he estimated it to be 180\,000 to 1.

{\sl All of the above, and especially the last two arguments,
  allows us to confidently state that the GRB~231115A burst
  was indeed not a short gamma-ray burst associated with the
  merger of neutron stars, but a giant flare of a previously
  unknown magnetar in the M\,82 galaxy.}

\section*{ACKNOWLEDGMENTS}
\noindent
The work was based on data from the INTEGRAL observatory,
obtained through its Russian and European Science Data Centers,
and the {\sl Fermi} observatory, obtained through
NASA/HEASARC. PM is grateful to the Time Allocation Committee
(TAC) of the INTEGRAL observatory for supporting his proposal
\#\,2040014 in response to the INTEGRAL Announcement of
Opportunity~20, which resulted in the data of GRB~231115A
observations.

\section*{FUNDING}
\noindent
The authors (AP, ICh, NP, PM, and SG) are grateful to the
Russian Science Foundation for financial support (grant
23-12-00220).\\ [2mm]

\begin{appendix}
\section*{ENERGY FLUX ACCORDING TO SPI-ACS}
\noindent
The INTEGRAL/SPI-ACS detector records the photon count rate in a
wide single energy channel of 0.085--10 MeV with a time
resolution of 50 ms. Since the studied GRB~231115A had a
duration of about 100 ms, its light curve based on the SPI-ACS
data consists in Fig.\,\ref{fig:ibis-lcurve} of only two
consecutive bins with a total significance of about 10 standard
deviations. It is noticeably inferior in terms of information
content to the light curves measured by other instruments.

At the same time, the SPI-ACS data allow us independently
estimate the energy flux from GRB~231115A using the results of
the SPI-ACS calibration, based on the analysis of joint
detection of the large number of GRBs in the INTEGRAL/SPI-ACS
and {\sl Fermi}/GBM experiments (Minaev \& Pozanenko 2023a). The
calibration included a study of the dependence of the SPI-ACS
effective area on the direction to the source vs the INTEGRAL
spacecraft's orientation and on the hardness of the emission
spectrum of the burst.

GRB~231115A was detected in the field of view of the INTEGRAL
aperture telescopes (the angle between the direction to the
source and the center of the field of view $z = 3\fdg8$), in
this case the effective area of SPI-ACS is close to the minimum
value.  According to Minaev \& Pozanenko (2023a) the conversion
factor of instrumental counts of SPI-ACS into energy units
erg~cm$^{-2}$ in the range of 10--1000 keV for a gamma-ray burst
with the spectral hardness $E_{p} = 640$ keV, which source has
coordinates in the detector coordinate system ($a$, $z$) =
(--109\fdg4, 3\fdg8), is $k = 5.1\times10^{-10}$
erg~cm$^{-2}$~count$^{-1}$. The time-integrated flux of
GRB~231115A according to the SPI-ACS detector is $F = 1117 \pm
107$ counts or $S = k\,F = (5.7_{-1.4}^{+1.7})\times 10^{-7}$
erg~cm$^{-2}$\ in the range of 10--1000 keV.  In the estimation
of the flux error, in addition to the statistical error of
measurements, the systematic error of the calibration method is
also taken into account (see details in Minaev \& Pozanenko
2023a). The obtained value within the $1\sigma$ agrees with the
flux measurement within the framework of the spectral analysis
of the {\sl Fermi}/GBM data $S = (7.25 \pm 0.46) \times 10^{-7}$
erg~cm$^{-2}$\ (see the corresponding section of this work).
\end{appendix}


\end{document}